\documentclass[lettersize,journal]{IEEEtran}
\usepackage{amsmath,amsfonts}
\usepackage{algorithmic}
\usepackage{algorithm}
\usepackage{array}
\usepackage{textcomp}
\usepackage{stfloats}
\usepackage{url}
\usepackage{verbatim}
\usepackage{graphicx}
\usepackage{cite}
\usepackage{enumitem}
\usepackage{subfigure}
\usepackage{multirow}
\usepackage{booktabs}

\usepackage{hyperref}

\usepackage{listings}
\usepackage{listings, xcolor}
\definecolor{verylightgray}{rgb}{.97,.97,.97}
\lstdefinelanguage{Solidity}{
  keywords=[1]{anonymous, assembly, assert, balance, break, call, callcode, case, catch, class, constant, continue, constructor, contract, debugger, default, delegatecall, delete, do, else, emit, event, experimental, export, external, false, finally, for, function, gas, if, implements, import, in, indexed, instanceof, interface, internal, is, length, library, log0, log1, log2, log3, log4, memory, modifier, new, payable, pragma, private, protected, public, pure, push, require, return, returns, revert, selfdestruct, solidity, storage, struct, suicide, super, switch, then, this, throw, transfer, true, try, typeof, using, value, view, while, with, addmod, ecrecover, keccak256, mulmod, ripemd160, sha256, sha3}, 
  keywordstyle=[1]\color{blue}\bfseries,
  keywords=[2]{address, bool, byte, bytes, bytes1, bytes2, bytes3, bytes4, bytes5, bytes6, bytes7, bytes8, bytes9, bytes10, bytes11, bytes12, bytes13, bytes14, bytes15, bytes16, bytes17, bytes18, bytes19, bytes20, bytes21, bytes22, bytes23, bytes24, bytes25, bytes26, bytes27, bytes28, bytes29, bytes30, bytes31, bytes32, enum, int, int8, int16, int24, int32, int40, int48, int56, int64, int72, int80, int88, int96, int104, int112, int120, int128, int136, int144, int152, int160, int168, int176, int184, int192, int200, int208, int216, int224, int232, int240, int248, int256, mapping, string, uint, uint8, uint16, uint24, uint32, uint40, uint48, uint56, uint64, uint72, uint80, uint88, uint96, uint104, uint112, uint120, uint128, uint136, uint144, uint152, uint160, uint168, uint176, uint184, uint192, uint200, uint208, uint216, uint224, uint232, uint240, uint248, uint256, var, void, ether, finney, szabo, wei, days, hours, minutes, seconds, weeks, years},  
  keywordstyle=[2]\color{teal}\bfseries,
  keywords=[3]{block, blockhash, coinbase, difficulty, gaslimit, number, timestamp, msg, data, gas, sender, sig, value, now, tx, gasprice, origin},  
  keywordstyle=[3]\color{violet}\bfseries,
  keywords=[4]{[1]},
  keywordstyle=[4]\color{blue}\bfseries,
  identifierstyle=\color{black},
  sensitive=false,
  comment=[l]{//},
  morecomment=[s]{/*}{*/},
  commentstyle=\color{gray}\ttfamily,
  stringstyle=\color{red}\ttfamily,
  morestring=[b]',
  morestring=[b]"
}
\def\myModel{{\sc ABCTracer}} 
\def\contrastiveModel{{\sc Connector}} 

\begin{document}

\title{Track and Trace: Automatically Uncovering Cross-chain Transactions in the Multi-blockchain Ecosystems}

\author{Dan Lin,~\IEEEmembership{Member,~IEEE}, Ziye Zheng, Jiajing Wu,~\IEEEmembership{Senior Member,~IEEE}, Jingjing Yang, Kaixin Lin, Huan Xiao, Bowen Song, Zibin Zheng,~\IEEEmembership{Fellow,~IEEE}

\thanks{The work described in this paper is supported by the National Key Research and Development Program of China (2023YFB2704700), the National Natural Science Foundation of China (623B2102, 62372485 and 62332004), the Natural Science Foundation of Guangdong Province (2023A1515011314), the Department of Education of Guangdong Province (2024ZDZX1001), the Fundamental Research Funds for the Central Universities of China (24lgqb018) and Shanghai Committee of Science and Technology, China (23511101000). \textit{(Corresponding author: Jiajing Wu.)}} 

\thanks{Z. Zheng, J. Yang and K. Lin are with the School of Computer Science and Engineering, Sun Yat-sen University, Guangzhou, 510006, China. D. Lin, J. Wu and Z. Zheng are with the School of Software Engineering, Sun Yat-sen University, Zhuhai 519082, China, and the GuangDong Engineering Technology Research Center of Blockchain. H. Xiao is with the School of Software Engineering, Guizhou University, Guiyang, 550025, China. Bowen Song is with the Ant Group, Hangzhou, China. (Email: wujiajing@mail.sysu.edu.cn)}}

\markboth{IEEE Transactions on Services Computing,~Vol.~xx, No.~x, xxx~2025}
{Zheng \MakeLowercase{\textit{et al.}}:Track and Trace: Automatically Uncovering Cross-chain Transactions in the Multi-Blockchain Ecosystems}

\IEEEpubid{0000--0000/00\$00.00~\copyright~2025 IEEE}

\maketitle

\begin{abstract}
Cross-chain technology enables seamless asset transfer and message-passing within decentralized finance (DeFi) ecosystems, facilitating multi-chain coexistence in the current blockchain environment. However, this development also raises security concerns, as malicious actors exploit cross-chain asset flows to conceal the provenance and destination of assets, thereby facilitating illegal activities such as money laundering. Consequently, the need for cross-chain transaction traceability has become increasingly urgent. Prior research on transaction traceability has predominantly focused on single-chain and centralized finance (CeFi) cross-chain scenarios, overlooking DeFi-specific considerations. This paper proposes \myModel, an automated, bi-directional cross-chain transaction tracing tool, specifically designed for DeFi ecosystems. By harnessing transaction event log mining and named entity recognition techniques, \myModel~automatically extracts explicit cross-chain cues. These cues are then combined with information retrieval techniques to encode implicit cues. \myModel~facilitates the autonomous learning of latent associated information and achieves bidirectional, generalized cross-chain transaction tracing. Our experiments on 12 mainstream cross-chain bridges demonstrate that \myModel~attains {91.75\%} bi-directional traceability (F1 metrics) with self-adaptive capability. Furthermore, we apply \myModel~to real-world cross-chain attack transactions and money laundering traceability, thereby bolstering the traceability and blockchain ecological security of DeFi bridging applications.
\end{abstract}


\section{Introduction}
\label{subsec:introduction}

The swift advancement of blockchain technology has garnered widespread attention and investment globally~\cite{YUAN2016BLOCKCHAIN}. Presently, the blockchain ecosystem comprises a multitude of chains. According to DeFiLlama~\cite{DEFILLAMA2023TOTAL}, as of  January 2025, a staggering 348 public blockchains have been documented. However, the inherent ``autonomy" of blockchain technology has led to a siloed environment, wherein assets and data between disparate chains are unable to interoperate~\cite{zheng2018blockchain}. Cross-chain bridge applications, which can be categorized into centralized finance (CeFi) and decentralized finance (DeFi) ~\cite{TrustSpectrum}, have been developed to facilitate asset and information exchange between different blockchains~\cite {swan2015blockchain,robinson2021survey}. While they enable seamless transfers, they also introduce new security challenges~\cite{ou2022overview, lee2023sok}.


Notably, as interconnections between blockchains have become increasingly prevalent, criminals have begun exploiting cross-chain bridge applications for illicit activities, such as money laundering. The Financial Action Task Force's (FATF) June 2022 report on virtual asset risks~\cite{FATFReport2022} highlighted the issue of money laundering through cross-chain transactions or ``chain hopping". Subsequently, in October 2022, a report by blockchain company Elliptic revealed that the threat of cross-chain crime had escalated, with up to {\$4.1} billion in illegally laundered funds funneled through services such as cross-chain bridges~\cite{EllipticReport2022}. As of 2023, this figure has surged to {\$7} billion~\cite{EllipticReport2023}.

\IEEEpubidadjcol
It is unequivocal that cross-chain crime has emerged as a pressing security concern. By leveraging cross-chain transactions as intermediaries, these crimes achieve more complex and clandestine financial flows compared to single-chain transactions. Consequently, addressing the traceability of cross-chain transactions is of paramount importance. Effective solutions must be capable of accurately identifying and matching the unique and consistent transactions executed by cross-chain bridges on both the source and destination chains while accommodating the bidirectional requirements of the traceability process. Specifically, for cross-chain money laundering and other illicit activities, forward traceability is essential, entailing the ability to trace back from the source to each branch. Conversely, for cross-chain attacks and other malicious behaviors, it is sometimes necessary to perform reverse traceability, tracing back from suspicious transactions on the destination chain to the corresponding transactions on the source chain. This two-way mechanism is crucial for ensuring the reliability and comprehensiveness of traceability.

In recent years, research on blockchain transaction traceability has primarily focused on single-chain scenarios, employing heuristic rules to address specific security concerns\cite{chen2019market,gao2020tracking,wu2021towards,wu2023know,xia2021trade,li2022ttagn,liu2024fishing,wu2020phishers,chen2018detecting,hu2023bert4eth,wang2022demystifying,weber2019anti,wu2023towards}. However, these methods are only applicable to single ledgers and cannot be directly applied to cross-chain scenarios. Currently, although some research has targeted CeFi cross-chain bridges~\cite{Yousaf2019Tracing,Zhang2022CLTracer}, these methods rely on centralized APIs, exhibiting lower accuracy and failing to generalize to DeFi cross-chain bridge scenarios. In contrast, DeFi cross-chain bridge transaction traceability poses the following challenges:
\begin{itemize}[leftmargin=*, itemsep=0.6pt]
    \item \textbf{C1: Multi-ledger isolation.} Unlike single-chain scenarios, cross-chain scenarios involve multiple physically isolated ledgers; unlike CeFi scenarios, DeFi scenarios lack a centrally managed cross-chain ledger.
    \item \textbf{C2: Heterogeneous cross-chain mechanisms.} Unlike single-chain scenarios, cross-chain scenarios require consideration of the diverse implementation mechanisms of different cross-chain bridges; in contrast to CeFi scenarios, DeFi scenarios lack a centralized institution providing a unified internal query interface for diverse cross-chain bridges.
    \item \textbf{C3: Complexity in identifying anomalous transactions.} The DeFi cross-chain bridge scenario introduces a new security concern, wherein cross-chain transactions' source-chain deposits and target-chain withdrawals typically adhere to specific business rules. Still, malicious or anomalous transactions may violate these rules. Existing approaches neglect these aggressive cross-chain transactions.
\end{itemize}

\contrastiveModel~\cite{lin2024connector} is the sole current work examining the traceability of DeFi bridge transactions. This tool analyzes the log of deposit transaction events on the source chain, extracts cross-chain key information via rule mapping, and subsequently correlates withdrawal transactions on the destination chain based on a predefined set of rules. However, this approach requires predefined rules, hindering automated learning. Furthermore, regarding traceability, \contrastiveModel~only supports unidirectional traceability from the source chain to the destination chain.

In this paper, we propose an \underline{\textbf{a}}utomated, \underline{\textbf{b}}i-directional \underline{\textbf{c}}ross-chain transaction tracking and \underline{\textbf{trac}}ing framework, dubbed \myModel, specifically designed for DeFi cross-chain scenarios. To address \textbf{C1}, we employ analysis of public transaction log events on the chain to extract cross-ledger information pertinent to the transactions, thereby facilitating effective associations between disparate ledgers. To tackle \textbf{C2}, we leverage named entity recognition techniques from natural language processing to automatically learn and extract explicit cues (i.e., bridge-agnostic association clues, as detailed in Section~\ref{subsec:methodology}) across various cross-chain mechanisms, thereby realizing a generalized automated tracing methodology. To overcome \textbf{C3}, we utilize information retrieval techniques to effectively learn implicit clues (i.e., bridge-unique association clues, as detailed in Section~\ref{subsec:methodology}), thereby developing a traceability approach that is decoupled from a priori business rules.

To assess the efficacy of \myModel, we conduct large-scale experiments on a real-world dataset comprising cross-chain transaction pairs from various open-source DeFi bridges. Experimental results on this dataset demonstrate that \myModel~achieves forward tracing, backward tracing, and bidirectional tracing with F1 scores of {94.92\%}, {89.58\%}, and {91.75\%}, respectively. Furthermore, we evaluate the effectiveness of \myModel~in practical application scenarios through cross-chain attack transaction cases and cross-chain money laundering cases. In summary, our primary contributions are as follows:
\begin{itemize}[leftmargin=*, itemsep=0.6pt]
    \item \textbf{Methodology.} We propose the first automated, bi-directional transaction tracing tool, \myModel\footnote{Available at \url{https://github.com/Connector-Tool/ABCTracer}}~, tailored to DeFi cross-chain scenarios, providing a more comprehensive cross-chain transaction analysis tool for the blockchain ecosystem where multiple chains coexist.
    \item \textbf{Evaluation.} Our method achieves bi-directional cross-chain transaction traceability with an F1 score of up to {91.75\%}. It also enables the automated learning and extraction of both explicit and implicit cues across chains. In addition, our method implements backward tracing and bi-directional tracing, which is not possible with the out-of-the-art method.
    \item \textbf{Application.} We apply \myModel~to real-world cross-chain attack transactions and cross-chain money laundering traceability scenarios, successfully identifying 20 pairs of cross-chain attack transactions and 10 pairs of cross-chain money laundering-related transactions. Based on the tracing results, we also performed pattern analysis on the identified cross-chain attack transactions.
\end{itemize}

The rest of the paper is organized as follows. Section~\ref{subsec:preliminary} introduces cross-chain bridges along with essential concepts. Section~\ref{subsec:definition} delineates and elaborates on the traceability issues associated with DeFi cross-chain transactions. Section~\ref{subsec:methodology} presents the \myModel~framework. Section~\ref{subsec:experiments} evaluates and implements our approach in both experimental and real-world cross-chain scenarios. Section~\ref{subsec:related_work} summarizes related work on transaction tracing. Finally, Section~\ref{subsec:conclusion} provides a concluding summary of this work.

\section{Preliminary}
\label{subsec:preliminary}

\subsection{Blockchain and Smart Contracts}
A blockchain is a peer-to-peer public ledger, while permissionless blockchains are independent systems that maintain their non-interconnectable ledgers~\cite{benisi2020blockchain}. Ethereum is the first blockchain to support smart contracts and operates on the Ethereum Virtual Machine (EVM). Ethereum accounts are categorized into two types: externally owned accounts (EOAs), which are controlled by users who hold private keys, and contract accounts, which are governed by the smart contract code deployed on the network. Smart contracts are executable code on the blockchain that is typically written in the Solidity programming language~\cite{fang2023beyond}. Upon compilation, they produce binary code (bytecode) and an Application Binary Interface (ABI), the latter of which documents all functions and events offered by the contract along with their parameter specifications.

\subsection{Transaction and Event Log}
Transactions can be classified into two categories: external transactions and internal transactions. External transactions are initiated by externally owned accounts (EOAs), whereas internal transactions are triggered by smart contracts. An individual external transaction can lead to multiple internal transactions, creating a complex web of interactions. Each transaction typically includes essential data such as the sender's address, the transaction amount, and event logs. During the execution of a transaction, a smart contract may trigger predefined events that log relevant information. A smart contract event is a specialized data structure designed specifically for recording events in the logs of the Ethereum Virtual Machine (EVM)~\cite{li2023understanding}.

\subsection{CeFi Bridge and DeFi Bridge}
A cross-chain bridge is a technical architecture that enables the interaction of assets and information across disparate blockchains. Based on distinct trust and verification models, cross-chain bridges can be categorized into centralized cross-chain bridges (CeFi bridges) and decentralized cross-chain bridges (DeFi bridges)~\cite{TrustSpectrum}. CeFi bridges primarily rely on EOA without on-chain code, which is analogous to centralized exchanges. In CeFi bridges, a centralized entity assumes custody of digital assets and transaction data, maintaining an internal ledger to record the cross-chain asset transfer process. In contrast, DeFi bridges are predominantly built upon DApps with smart contracts, similar to decentralized exchanges. The on-chain router contract is responsible for interacting with users and token contracts, providing on-chain functionality, including token locking and unlocking, and recording token transfer information as on-chain events. Off-chain repeaters, in turn, are tasked with retrieving on-chain events and coordinating with router contracts on the target chain to facilitate cross-chain asset transfers.

\subsection{Named Entity Recognition and Information Retrieval}
Named Entity Recognition (NER)~\cite{jehangir2023survey} and Information Retrieval (IR)~\cite{hambarde2023information} are two pivotal tasks in the realm of Natural Language Processing (NLP). The primary objective of NER is to automatically identify and categorize diverse entities, including person names, geographic locations, and organizational entities, from unstructured text~\cite{jehangir2023survey}. In contrast, Information Retrieval (IR) entails extracting relevant data from a vast corpus of information in response to a user's query, typically relying on keyword-based indexing and ranking algorithms~\cite{hambarde2023information}.

\section{Problem Definition}
\label{subsec:definition}
After executing a transaction via a cross-chain bridge, directly obtaining the corresponding transaction on another chain is often not feasible. Similarly, retrieving the source chain transaction from the destination chain is also challenging. To address this issue, cross-chain transaction tracing techniques can be employed to track and trace transactions across different chains\footnote{For brevity, unless otherwise noted, ``cross-chain transaction tracing" in the following sections refers to both forward and backward tracing, i.e., cross-chain transaction tracking and tracing.}. This paper explores cross-chain transaction tracing in DeFi environments, aiming to uncover complete transaction pairs, specifically source chain deposits and destination chain withdrawals.

\noindent{\sc \textbf{Problem} (Cross-chain transaction tracing):}
\textit{Given a set of blockchain transactions comprising $N$ non-cross-chain transactions and $M$ query cross-chain transactions, the $M$ query cross-chain transactions involve $K$ distinct chains. Each transaction is uniquely identified by its transaction hash.}

\noindent$\bullet$ \textit{Identify the subset of query cross-chain transactions within the set $TX_{Q} = \{tx^{c_1}_1,\dots,tx^{c_i}_i,\dots,tx^{c_M}_M\},c_{i} \in [1,K]$;}

\noindent$\bullet$ \textit{Track and trace the corresponding target cross-chain transactions from K blockchains based on $TX_{Q}$, forming the target cross-chain transactions within the set $TX_{T} = \{tx^{c_1}_1,\dots,tx^{c_j}_j,\dots,tx^{c_M}_M\},c_{j} \in [1, K]$;}

\noindent$\bullet$ \textit{Obtain a set of cross-chain transaction pairs exhibiting consistent deposit and withdrawal behavior, represented as $\{(tx^{c_{i}}_{i},tx^{c_{j}}_{j})|tx^{c_{i}}_{i} \in TX_{Q},tx^{c_{j}}_{j} \in TX_{T}\}$.}

For clarity in subsequent descriptions, we denote the source and destination chains in a cross-chain transaction as ``Source'' and ``Destination'', respectively. Additionally, we refer to the chains of the to-be-queried transactions and the resulting paired transactions as ``Query'' and ``Target''. The corresponding transaction is represented as $tx^{S}$, $tx^{D}$, $tx^{Q}$ and $tx^{T}$. For instance, in a cross-chain transaction $(\href{https://etherscan.io/tx/0x2f13d202c301c8c1787469310a2671c8b57837eb7a8a768df857cbc7b3ea32d8#eventlog}{0x2f13d},\href{https://bscscan.com/tx/0xfd60c2ce27c4f58f8020918c0a20dbabc2a55f794dc7d657f8c1173e8e2f1d58#eventlog}{0xfd60c})$ from Ethereum to Binance Smart Chain, Ethereum is the ``Source'' and Binance Smart Chain is the ``Destination''. In forward tracing task, Ethereum serves as the ``Query'' and Binance Smart Chain as the ``Target''. Conversely, in backward tracing, Binance Smart Chain is the ``Query'', while Ethereum remains the ``Target''.

\section{Methodology}
\label{subsec:methodology}
\begin{figure}[h]
\setlength{\abovecaptionskip}{0.2cm}
    \centering
    \includegraphics[width=1\linewidth]{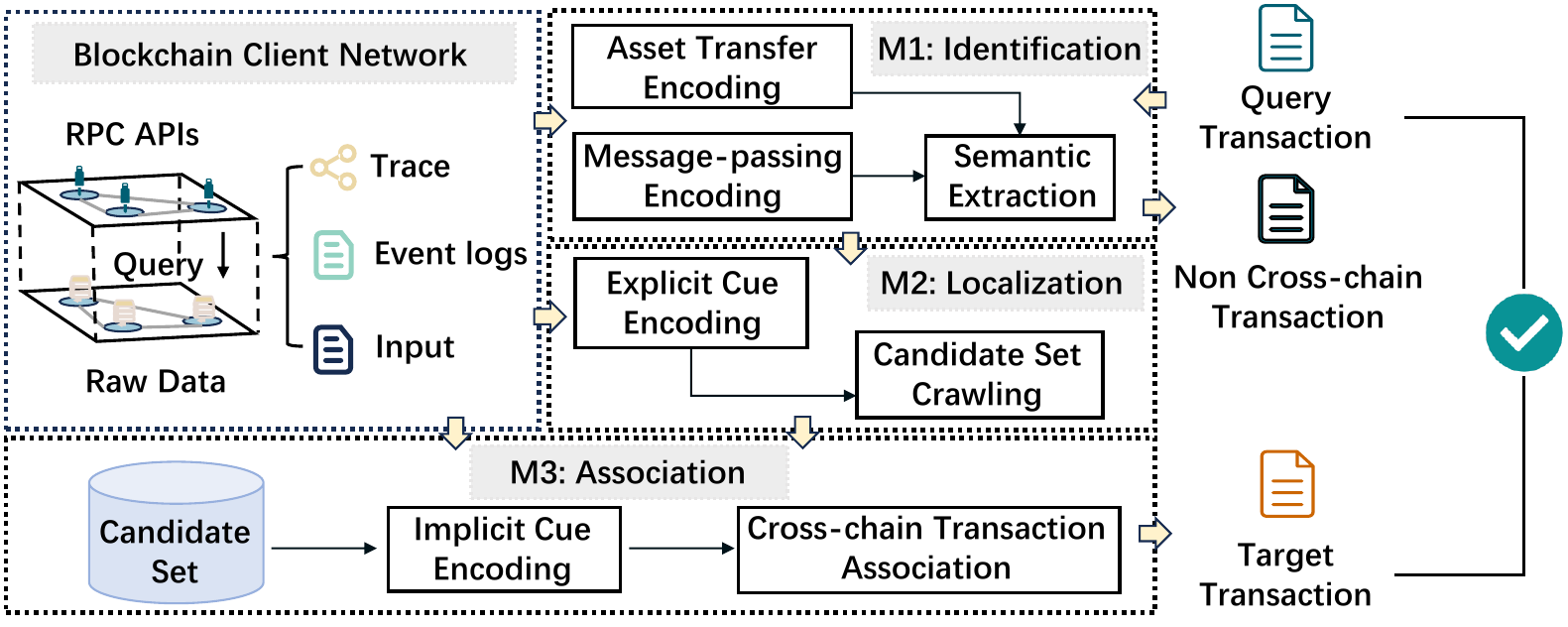}
    \caption{Overview of \myModel, including three main modules (\textbf{M1}: cross-chain transaction identification based on semantic extraction, \textbf{M2}: candidate transactions localization based on explicit clues, \textbf{M3}: cross-chain transaction association based on implicit clues).}
    \label{fig:framework}
\end{figure}

We design an automated, bi-directional cross-chain transaction tracing framework, \myModel, tailored for DeFi scenarios, as illustrated in Fig.~\ref{fig:framework}. \myModel~is capable of autonomously learning and extracting key cross-chain information within DeFi scenarios, supporting bi-directional traceability from the source chain to the destination chain and vice versa. \myModel~takes transaction information—including timestamps, transaction inputs, and event logs as inputs and outputs complete cross-chain transaction pairs. Specifically, \myModel~comprises three main modules:
\begin{itemize}[leftmargin=*, itemsep=0.6pt]
    \item \textbf{M1: cross-chain transaction identification based on semantic extraction.} Employs statistical and linguistic modeling to mine asset transfer and message-passing semantics from the transaction trace and event logs, determining whether the query transaction qualifies as a cross-chain transaction.
    \item \textbf{M2: candidate transactions localization based on explicit clues.} Utilizes NER to automatically learn explicit cross-chain clues from the query transaction inputs and event logs, using these clues as constraints to crawl the set of candidate target transactions.
    \item \textbf{M3: cross-chain transaction association based on implicit clues.} Leverages IR to identify implicit clues between query transactions and candidate target transactions, enabling precise associations between the query and target transactions. Ultimately, produces complete cross-chain transaction pairs.
\end{itemize}

\subsection{\textbf{M1}: Cross-chain Transaction Identification}
In cross-chain bridge applications, the first step for a user to conduct a cross-chain transaction is usually to initiate a deposit transaction on the source chain. Subsequently, messages are transmitted through an off-chain relay, ultimately triggering a withdrawal transaction on the destination chain, thereby completing the cross-chain operation. Unlike non-cross-chain transactions, the integrity of a cross-chain transaction is defined by the combination of the deposit on the source chain and the withdrawal on the destination chain. This cross-chain characteristic necessitates the traceability of transactions. Therefore, to effectively identify cross-chain transactions, we develop a cross-chain semantic extraction module designed to differentiate cross-chain transactions from non-cross-chain transactions. This approach narrows the scope of traceability and enhances both the effectiveness and efficiency of the overall tracing process.

\begin{figure}[h]
\setlength{\abovecaptionskip}{0.2cm}
    \centering
    \includegraphics[width=1\linewidth]{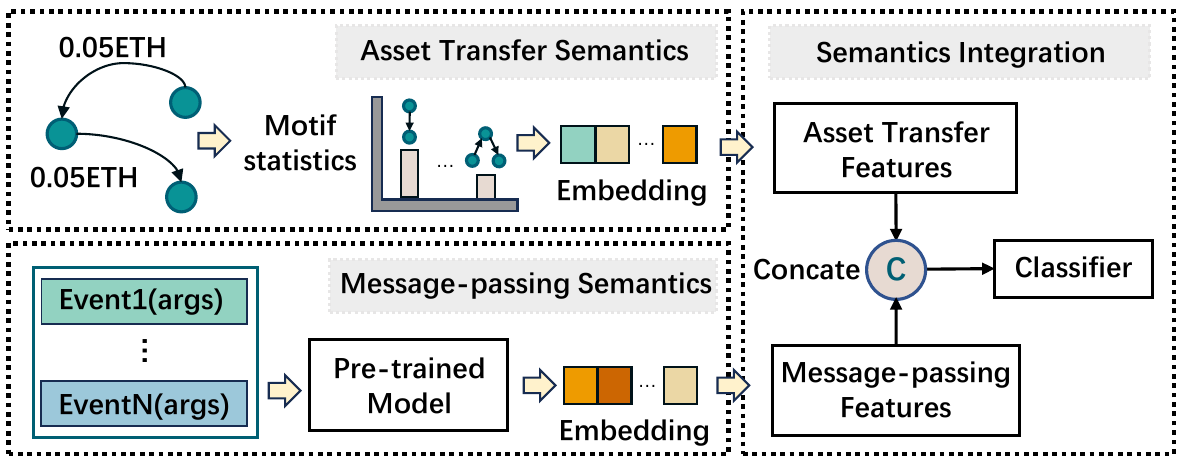}
    \caption{Cross-chain transaction identification model that extracts and integrates asset transfer and message-passing semantics, followed by a classifier to distinguish cross-chain from non-cross-chain transactions.}
    \label{fig:m1}
\end{figure}

In this module, we employ XSema~\cite{zheng2024xsema} to facilitate cross-chain transaction identification, as illustrated in  Fig.~\ref{fig:m1}. Our proposed XSema~\cite{zheng2024xsema} offers a comprehensive analysis of cross-chain semantics, identifying two key elements as its semantic sources: asset transfer and message-passing. First, we utilize statistical modeling techniques to encode the transaction asset transfer graph, generating a corresponding 16-dimensional semantic representation for asset transfer. Next, we encode the event name text within the transaction event logs using a pre-trained text model, followed by densifying the encoded results with a multilayer perceptron to obtain a 16-dimensional semantic representation for message-passing. Subsequently, we construct the final semantic representation of the transaction by integrating the asset transfer semantics with the message-passing semantics. Finally, we we utilize a classifier to accurately identify and output cross-chain transactions, including deposit transactions on source chains and withdrawal transactions on destination chains.

\subsection{\textbf{M2}: Candidate Transactions Localization}
Typically, deposit and withdrawal transactions triggered on cross-chain bridges retain certain associative clues in both temporal and spatial dimensions. In this paper, we refer to these common clues as explicit clues. In the temporal dimension, the withdrawal transaction on the destination chain is usually initiated sometime after the deposit transaction on the source chain, making the time interval a significant temporal cue in the tracing process~\cite{lin2024connector}. In the spatial dimension, cross-chain activities involve both parties across the chains, including chain and address information. Therefore, the destination chain and corresponding destination address serve as important spatial clues during tracing.

This module automatically learns and extracts explicit clues to derive a set of candidate transactions associated with the query transaction. By comprehensively considering the constraints provided by explicit clues, we can concentrate more effectively on the tracing target and narrow the scope of the trace, thereby enhancing both the efficiency and accuracy of the tracing process.

However, as mentioned in Section~\ref{subsec:introduction}, the physical isolation of inter-chain ledgers (\textbf{C1}) and the diversity of cross-chain mechanisms (\textbf{C2}) present challenges in acquiring spatial clues. To address \textbf{C1}, we leverage the message-passing mechanism of cross-chain bridges to mine the event log sequences of cross-chain transactions, thereby deriving logical association information between chains. For \textbf{C2}, we utilize named entity recognition technology to automatically learn and extract entities related to target chains and addresses, facilitating the acquisition of spatial clues. Ultimately, by integrating both temporal and spatial clues, we can narrow the tracing scope and ensure a solid foundation for subsequent associations. Specifically, this module can be divided into three distinct steps: 1) acquisition of temporal clues, 2) acquisition of spatial clues, and 3) crawling for candidate target transactions.

\subsubsection{Acquisition of Temporal Clues} According to the cross-chain transaction business logic specified by the cross-chain bridge, withdrawal transactions are typically completed within 30 minutes (or less) following the confirmation of the deposit transaction. Therefore, we establish an optimal time interval constraint for each cross-chain bridge application in advance (for details, see Section~\ref{subsec:experiments}), which will serve as a temporal cue in the tracing process.

\subsubsection{Acquisition of Spatial Clues} Based on the description of the cross-chain message-passing mechanism provided on the official cross-chain bridge website and our observations of transaction event log texts, we find that in most cases, logs related to cross-chain operations contain event names that include information about the target chain and target address. Fig.~\ref{fig:event} illustrates the event declaration for the withdrawal transaction $\href{https://polygonscan.com/tx/0x6545620535502aedd8c40843c8c28facddf403e2fa2729070afd44848d206a52}{0x65456}$ on the Polygon PoS Chain, where the \texttt{originChainId} formal parameter corresponds to the source chain, and the \texttt{depositor} formal parameter corresponds to the address of sender. The event also includes details about the \texttt{destinationChainId}, which indicates the chain where the transaction is queried, as well as information about the address of recipient labeled as \texttt{recipient}. Consequently, the trail object can be identified based on the names of the formal parameters.

\begin{figure}[tbp]
\setlength{\abovecaptionskip}{0.2cm}
\begin{lstlisting}[numbers=none]
FundsDeposited (uint256 amount, uint256 originChainId, uint256 destinationChainId, uint64 relayerFeePct, index_topic_1 uint32 depositId, uint32 quoteTimestamp, index_topic_2 address originToken, address recipient, index_topic_3 address depositor)
\end{lstlisting}
\caption{Example event declaration for withdrawal transaction. }
\label{fig:event}
\vskip -3ex
\end{figure}

\begin{figure}[h]
\setlength{\abovecaptionskip}{0.2cm}
    \centering
    \includegraphics[width=0.8\linewidth]{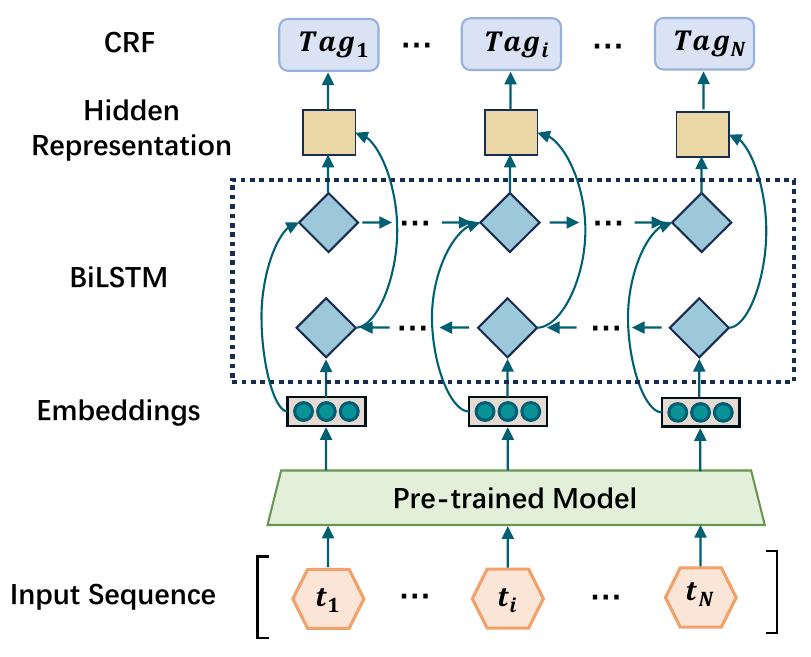}
    \caption{Overview of the candidate transactions localization model. Input sequences are processed through a pre-trained model to generate embeddings, which are then used in a BiLSTM for contextual learning and passed to a CRF for final token tagging.}
    \label{fig:m2}
\end{figure}

It is important to note that variations in cross-chain mechanisms lead to differences in the naming of formal parameters associated with these clues. As a result, existing methods struggle to adapt effectively to newly emerging cross-chain bridges, necessitating the manual definition of expert rules for parameter matching. To address this, we introduce named entity recognition technique to learn the naming conventions of formal parameters and automatically extract the corresponding parameter values as spatial clues. we utilize the currently mainstream pre-trained model + BiLSTM + CRF~\cite{li2020survey} as the framework for entity extraction, as shown in Fig.~\ref{fig:m2}.

Specifically, considering that the function texts in the event name and transaction input are analogous to function declarations in code, we adopt a code text pre-training model to encode each word in the text, thereby generating the corresponding embedding representations. Subsequently, we employ a BiLSTM~\cite{graves2012long} network to learn contextual information, enhancing entity recognition by taking into account both forward and backward information within the sequences. Building on this, we further introduce Conditional Random Fields (CRF)~\cite{wallach2004conditional} to globally optimize the features extracted by the BiLSTM, enhancing annotation accuracy by modeling the dependencies between labels in the sequence labeling task. For instance, each token $t_{i}$ in the input sequence will ultimately receive a corresponding $Tag_{i}$, representing the entity type of that token.

\subsubsection{Crawling for Candidate Target Transactions} After establishing the time interval, target chain, and target address, we utilize the BlockchainSpider~\cite{wu2023tracer} to crawl all transactions associated with the target address on the target chain within the specified time interval. This process localizes a set of candidate target transactions, effectively narrowing the scope of traceability.

Through the localization provided by this module, we obtain a set of candidate transactions related to the query transaction on the target chain. Compared to directly retrieving all transactions from time-adjacent blocks on the target chain, this approach significantly reduces the search space, thereby establishing a solid foundation for the precise traceability of \textbf{M3}.

\subsection{\textbf{M3}: Cross-chain Transaction Association}

Through \textbf{M2}, we can retrieve a set of candidate target transactions that exhibit a certain correlation with query transaction, derived from extracting explicit clues during the tracing process. The basis of association needs to be reinforced to further achieve precise matching with the ultimate target transactions. The transaction amount-cost ratio is an intuitive and effective choice. According to the business logic of cross-chain transactions, the cross-chain transaction fee typically ranges from {0\%} to {3\%} of the deposit transaction amount~\cite{CelerFAQ2}. However, in actual cross-chain scenarios, malicious or abnormal transactions may violate this rule (\textbf{C3}). To mitigate \textbf{C3}, we discover implicit cues. Specifically, implicit cues refer to the unique identification patterns inherent to each cross-chain bridge. These cues serve as the foundation for associations within the cross-chain mechanism, relying on intelligent learning tools to uncover important information that experts may not be able to identify based on prior knowledge. This contrasts with explicit cues, which experts can directly recognize as significant.

\begin{figure}[h]
\setlength{\abovecaptionskip}{0.2cm}
    \centering
    \includegraphics[width=1\linewidth]{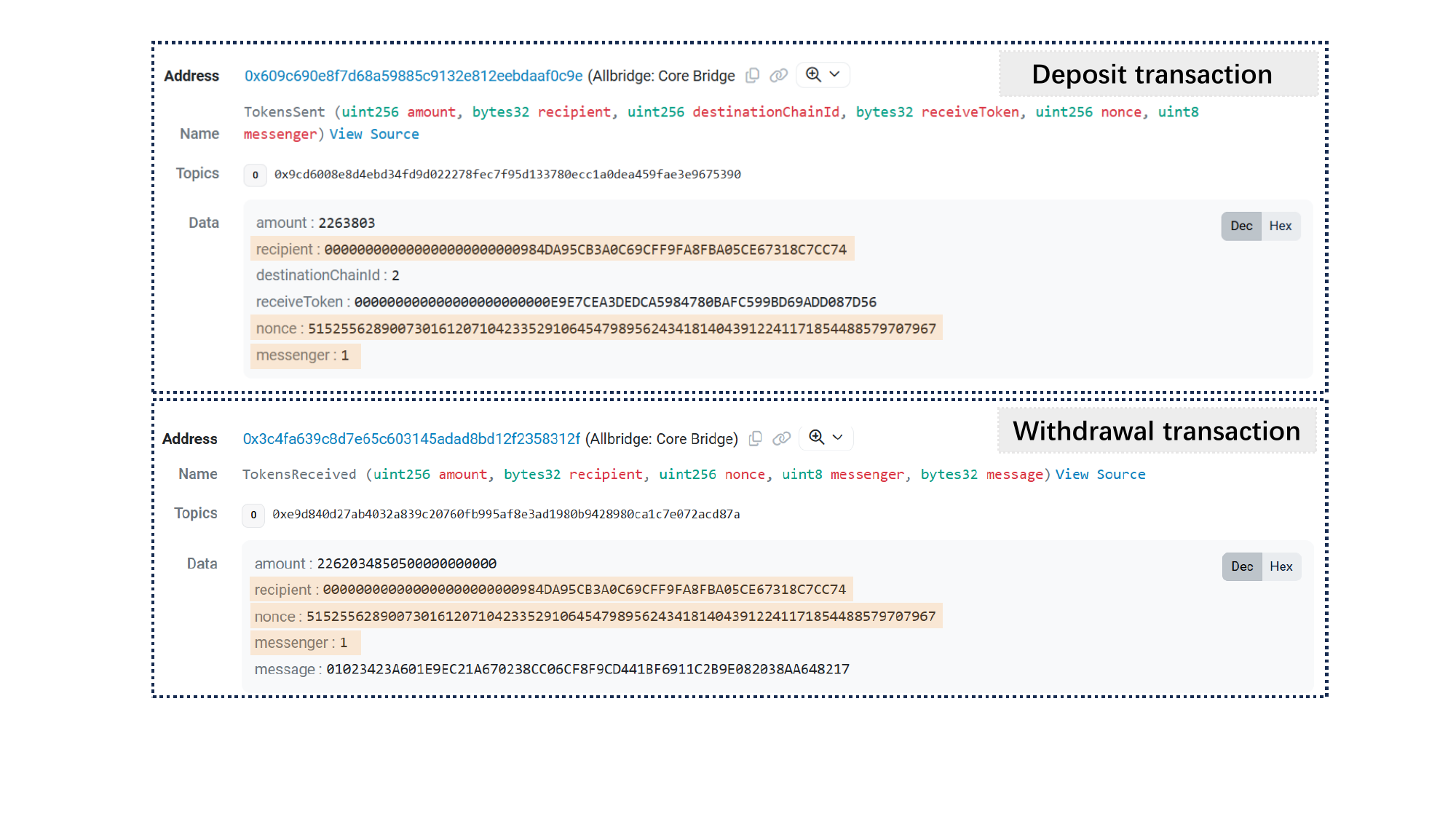}
    \caption{An example of an Allbridge cross-chain transaction from Ethereum to Binance Smart Chain, with implicit cues including the \texttt{recipient}, \texttt{nonce}, and \texttt{messenger} parameters.}
    \label{fig:implicit_clue}
\end{figure}

\begin{figure}[h]
\setlength{\abovecaptionskip}{0.2cm}
    \centering
    \includegraphics[width=1\linewidth]{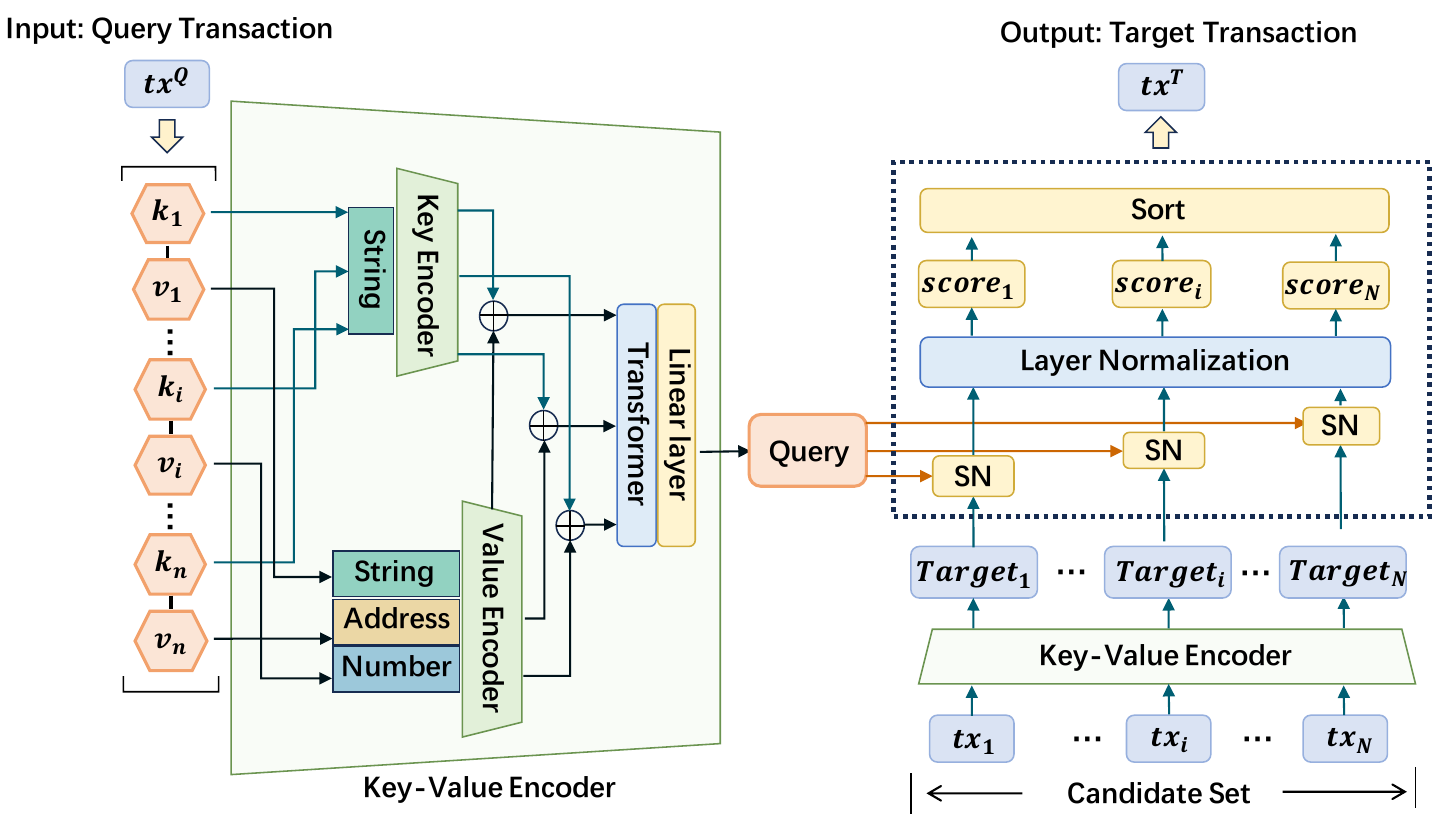}
    \caption{Cross-chain transaction association model with implicit cue encoding and transaction association processes input query transactions to yield the target transaction exhibiting the highest association.}
    \label{fig:m3}
\end{figure}

To demonstrate the Allbridge-specific implicit cues, an example of an Allbridge cross-chain transaction $(\href{https://etherscan.io/tx/0x2f13d202c301c8c1787469310a2671c8b57837eb7a8a768df857cbc7b3ea32d8#eventlog}{0x2f13d},\href{https://bscscan.com/tx/0xfd60c2ce27c4f58f8020918c0a20dbabc2a55f794dc7d657f8c1173e8e2f1d58#eventlog}{0xfd60c})$ from Ethereum to Binance Smart Chain is shown in Fig.~\ref{fig:implicit_clue}. The cross-chain log of the deposit transaction 0x2f13d contains the \texttt{recipient}, \texttt{nonce}, and \texttt{messenger} parameters, whose values are consistent with those of the \texttt{recipient}, \texttt{nonce}, and \texttt{messenger} parameters in the withdrawal transaction $\href{https://bscscan.com/tx/0xfd60c2ce27c4f58f8020918c0a20dbabc2a55f794dc7d657f8c1173e8e2f1d58#eventlog}{0xfd60c}$, respectively. This implies that the cross-chain bridge preserves associative cues between the deposit transaction and the withdrawal transaction. Moreover, these cues typically vary across different cross-chain bridges. Based on the observation of potential implicit cues, we opt to encode key-value pair information in the event log and employ information retrieval techniques~\cite{pang2020setrank} to establish transaction associations. This approach not only effectively preserves the business logic of the transaction amount and cost ratio through the amount field in the key-value pair, but also combines the unique implicit clues of the cross-chain bridge to effectively enhance the association effect. Specifically, this module is shown in Fig.~\ref{fig:m3} which can be further subdivided into two stages: 1) implicit cue encoding and 2) cross-chain transaction association.

\subsubsection{Implicit Cue Encoding} Function and event parameters capture the distinctive behavioral attributes of each transaction. Consequently, by encoding the parameter dictionary to represent the transaction $tx^{Q}$ itself, latent implicit cues can be effectively extracted, thereby further enhancing the transaction association effect. Noting that implicit clues typically manifest in the parameter dictionary as string, address, and numeric types, this paper identifies key-value pairs with values belonging to these three types as encoding objects:

\begin{itemize}[leftmargin=*, itemsep=0.6pt]
    \item \textbf{Numeric types.} Convert numeric data to floating point numbers and get the corresponding binary sequence. Convert the binary sequence to a vector, where each dimension of the vector is represented by a 0 or a 1.
    \item \textbf{String types.} Obtain corresponding embedding representations using pre-trained code language models such as CodeBert~\cite{feng2020codebert}.
    \item \textbf{Address types.} Create learnable embedded representations for address types, similar to Bert4ETH~\cite{hu2023bert4eth}.
\end{itemize}

Key-value pairs (i.e., {$(k_1, v_1), ...,(k_n,v_n)$}) in these three types are encoded and then fed into separate MLP layers to obtain a dense representation with consistent dimensionality. Specifically, when encoding the key-value pairs in the parameter dictionary, we employ string-based encoding, considering that the keys are strings. The values are then encoded according to their respective types. Subsequently, the obtained key vectors are concatenated with the value vectors to form a comprehensive key-value pair representation. All key-value pairs corresponding to the transaction are then input into the Transformer layer to learn the associative relationships between each key-value pair. Ultimately, the embedded representation incorporating complete key-value pair information is fed into the Linear layer to generate the final embedded representation of the transaction.

\subsubsection{Cross-chain Transaction Association} To calculate the degree of association between the query transaction and each transaction in the set of candidate target transactions, we feed the implicit cue encodings of the transactions into a Learning Sorting Network and select the transaction with the highest association as the final correlated object based on the sorting outcome. Specifically, the embedding representations corresponding to the query transaction and the embedding representations corresponding to the $m$ candidate target transactions are input into a Siamese Network (SN) architecture, which yields the association degree of each transaction, respectively:

\begin{equation}
\label{eq:sn}
\begin{gathered}
P=\left\{p_1, p_2, \ldots, p_N\right\} \\
p_i=g\left(\left[f\left(q ; \theta_{s n}\right) ; f\left(d_i ; \theta_{s n}\right)\right] ; \varphi\right), i \in[1, N]
\end{gathered}
\end{equation}
where, $f\left(\cdot ; \theta_{s n}\right)$ denotes a multilayer neural network of SN, $\theta_{s n}$ denotes the network parameters, $g\left(\cdot ; \varphi\right)$ denotes a fully connected layer, and $\varphi$ denotes the layer's parameters. Subsequently, the obtained correlated feature representation is fed into a Layer Normalization (Layer Norm) layer to yield the normalized representation:

\begin{equation}
\label{eq:norm}
\begin{gathered}
\begin{gathered}
\hat{P}=\left\{\hat{p}_1, \hat{p}_2, \ldots, \hat{p}_N\right\} \\
score_{i} = \hat{p}_i=\gamma\left(\frac{p_i-\mu_i}{\sqrt{\sigma_i^2+\epsilon}}\right)+\beta, i \in[1, N]
\end{gathered}
\end{gathered}
\end{equation}

The output of each Layer Norm layer is utilized as the association score corresponding to the respective transaction. As depicted in Eq.~\ref{eq:norm}, $\mu_i$ represents the vector mean of $p_i$, $\sigma_i^2$ denotes its variance, while $\gamma$ and $\beta$ signify the scaling and translation parameters learned by the Layer Norm layer. Furthermore, $\epsilon$ is an infinitesimally small constant introduced to prevent division by zero, and $\hat{p}_i$ denotes the normalized value. Ultimately, the candidate transactions are ranked by their association scores, and the transaction exhibiting the highest association is selected as the target transaction.

Based on the set of candidate transactions obtained by \textbf{M2}, this module encodes the candidate transactions with implicit clues and calculates the degree of association between the query transaction and each candidate transaction. Ultimately, we use the sorting method to identify the target transaction with the highest degree of association as the final output, thus realizing cross-chain traceability between the deposit transaction of the source chain and the withdrawal transaction of the destination chain.

\section{Experiments}
\label{subsec:experiments}

In this section, we conduct evaluations to demonstrate the validity of the proposed \myModel~framework. Specifically, we aim to answer the following research questions (RQs):
\begin{itemize}[leftmargin=*, itemsep=0.6pt]
    \item \textbf{RQ1: Model effectiveness.} How effective is the proposed \myModel~framework for tracing cross-chain transactions in DeFi scenarios bi-directionally and automatically?
    \item \textbf{RQ2: Application to attack transactions.} How effective is the proposed \myModel~framework in tracing anomalous cross-chain transactions and identifying attack transactions in the real-world application of our approach?
    \item \textbf{RQ3: Application to money laundering transactions.} How effective is our approach in tracing real-world cross-chain money laundering cases?
\end{itemize}

\subsection{Datasets}
In this experiment, we select representative bridges based on the following criteria and collect relevant contract and transaction data as the primary source for the experiment: 1) bridges with the highest liquidity in the first quarter of 2023~\cite{Liquidity}; 2) having a bridge browser service that can be used to build real datasets; 3) supporting EVM-compatible blockchains; and 4) experiencing security incidents. Based on the above criteria, 12 cross-chain bridges are finally selected, namely, Celer cBridge, Multichain, Poly Network, Allbridge, Connext, Debridge, Stargate, Symbiosis, Synapse protocol, Transit Swap, Wormhole, and Router protocol. It is worth noting that the \myModel~discussed in this study applies to other contract-based, non-privacy-preserving, EVM-compatible DeFi bridges.

The dataset we used is consistent with the dataset mentioned in \contrastiveModel~\cite{lin2024connector} to facilitate comparison experiments. Specifically, this dataset is a contract and transaction dataset involving 12 cross-chain bridges. Given that Ethereum is the largest permissionless blockchain platform supporting smart contracts, this dataset using Ethereum as the source chain. It obtains the set of Ethereum contract addresses from the official website of the bridge and collects contract transactions from April 2021 to March 2024. Then, it categorizes and labeles the collected transactions based on the query results of the cross-chain bridge browser, extracting the cross-chain transactions therein. Following this, it performs a statistical analysis of the blockchains with the highest co-transaction volumes. The results indicate that the top three blockchains are Ethereum (ETH), Binance Smart Chain (BSC), and Polygon PoS Chain (MATIC). In summary, this dataset extracts 29,289 real cross-chain transaction pairs, comprising 24,926 cross-chain transaction pairs initiated from ETH and received by BSC, and 4,363 cross-chain transaction pairs initiated from ETH and received by MATIC. Therefore, we will focus on evaluating cross-chain transaction tracing involving ETH, BSC, and MATIC.

\subsection{Baselines and variants}
To scientifically assess the performance of \myModel~in cross-chain transaction tracing, we adopt \contrastiveModel~\cite{lin2024connector} as the benchmark methodology. Moreover, we design a variant of the pre-training module within \myModel, leveraging a diverse set of four pre-training models, including CodeBert~\cite{feng2020codebert}, GraphCodeBert~\cite{guo2020graphcodebert}, UnixCoder~\cite{guo2022unixcoder}, and CodeT5~\cite{wang2021codet5}, to comprehensively evaluate the generalizability and robustness of \myModel.

\subsection{RQ1: Model effectiveness}
To address RQ1, we evaluate the performance of \myModel~in tracing cross-chain transactions within the DeFi context. We conduct experiments involving bi-directional transaction tracing and automated learning to systematically examine the framework's effectiveness in bidirectional tracing of cross-chain transactions, as well as its capability to autonomously learn cross-chain cues.

\subsubsection{Capability of Bidirectional Tracing} Based on the cross-chain transaction dataset with real labels, we conduct forward tracing, backward tracing, and bi-directional tracing of cross-chain transactions, respectively. 
Specifically, in the forward tracing experiment, we trace the deposit transactions obtained for M1, that is, tracing from the source chain deposit transaction to the target chain withdrawal transaction. In the backward tracing experiment, we trace the withdrawal transactions obtained for M1, that is, tracing from the target chain withdrawal transaction to the source chain deposit transaction. In the bidirectional tracing experiment, we simultaneously perform track and trace on all cross-chain transactions (i.e., deposits and withdrawals) obtained for M1.
To mitigate the potential impact of uneven event log distributions on our experiments, we count the number of event logs associated with each transaction to be traced and subsequently distribute the data to the training, validation, and test sets in a balanced manner. Concurrently, we divided the training set, validation set, and test set in a 7:1.5:1.5 ratio and determined suitable time interval parameters for each cross-chain bridge.

\begin{figure}[t]
\setlength{\abovecaptionskip}{0.2cm}
    \centering
    \hspace{-2.5ex}
	\subfigure[Celer cBridge]{%
        \centering
        \includegraphics[width=0.32\linewidth]{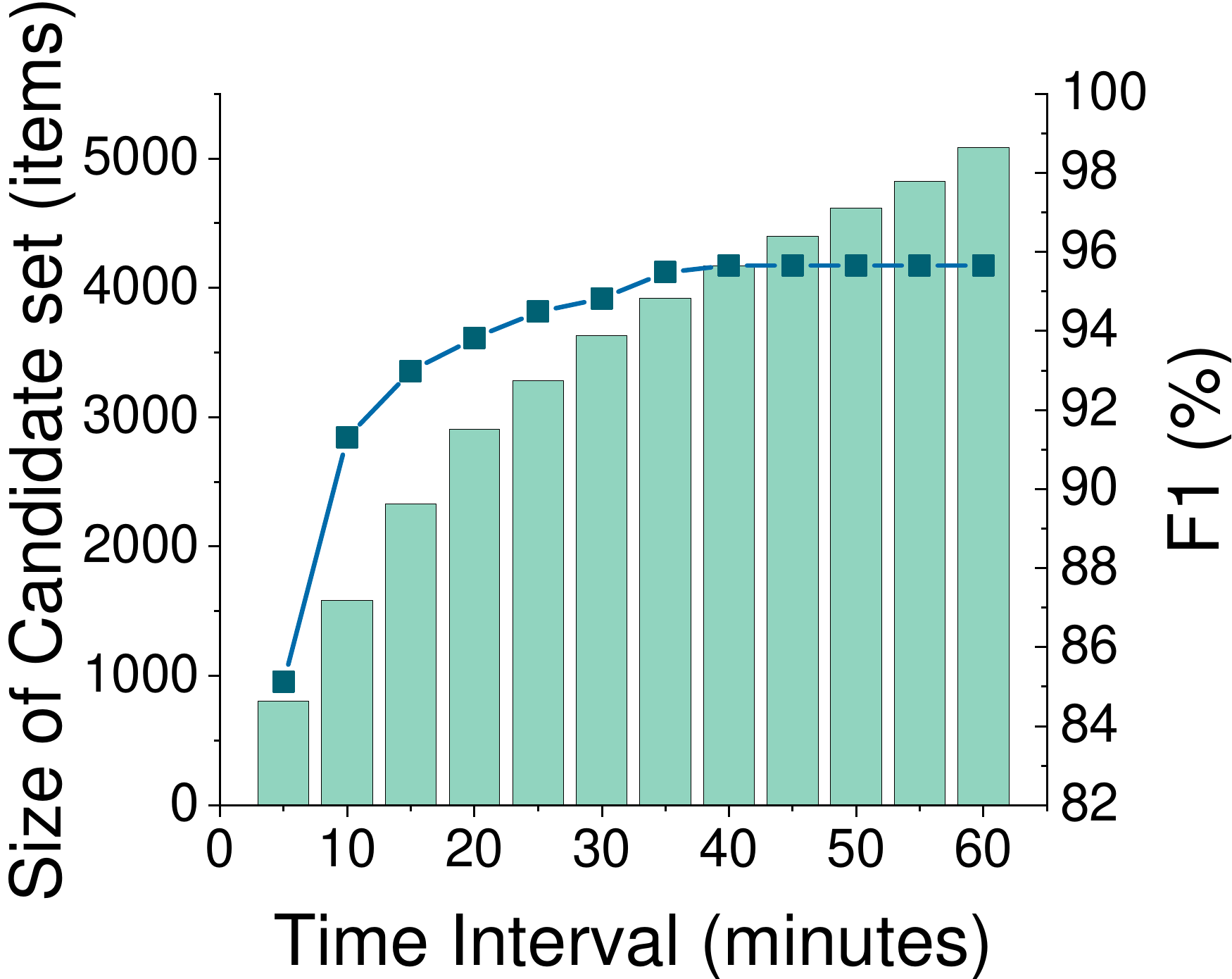}
	}
    \hspace{-0.6ex}
    \subfigure[Multichain]{%
        \centering
        \includegraphics[width=0.32\linewidth]{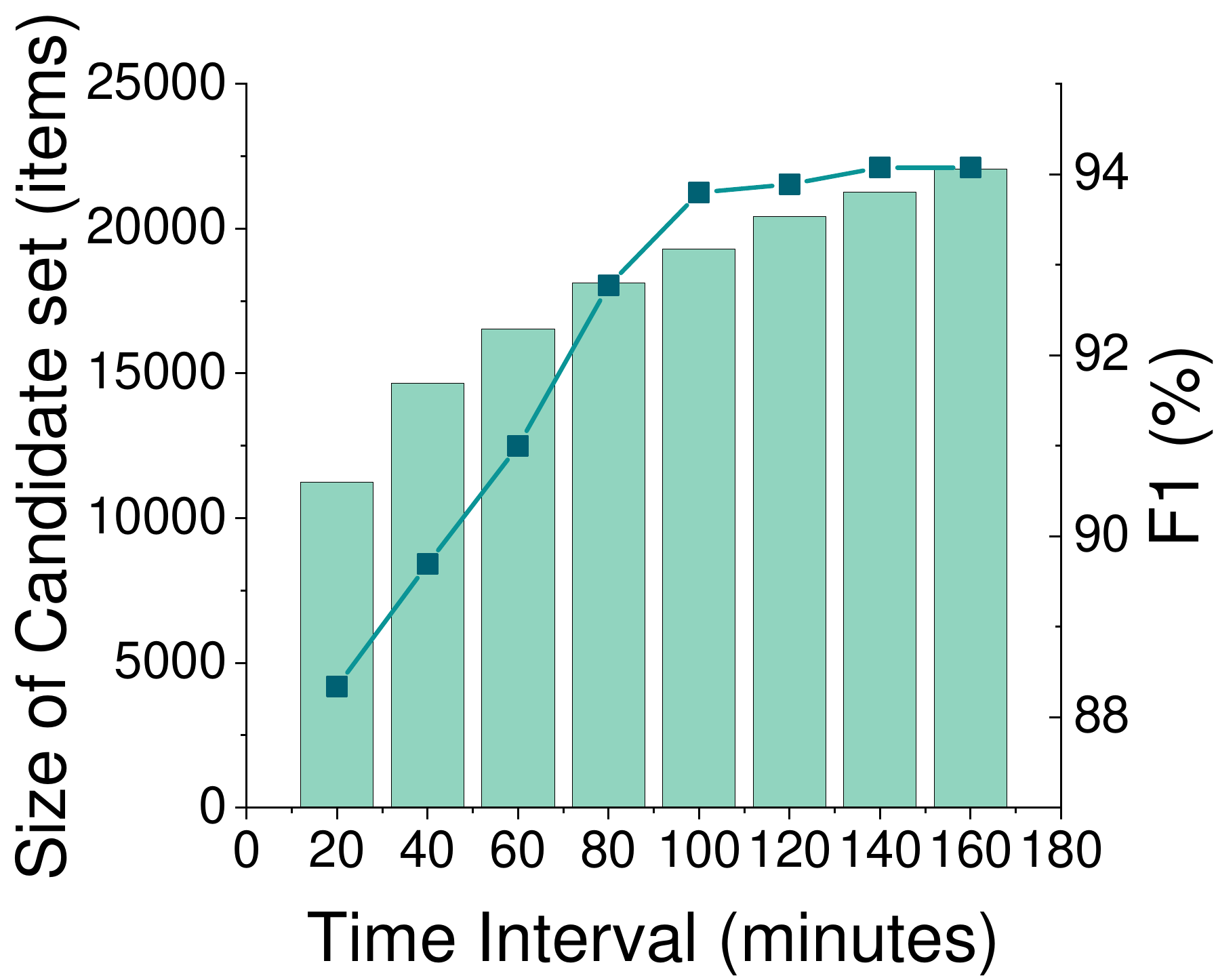}
	}    
    \hspace{-0.6ex}
	\subfigure[Poly Network]{%
        \centering
        \includegraphics[width=0.32\linewidth]{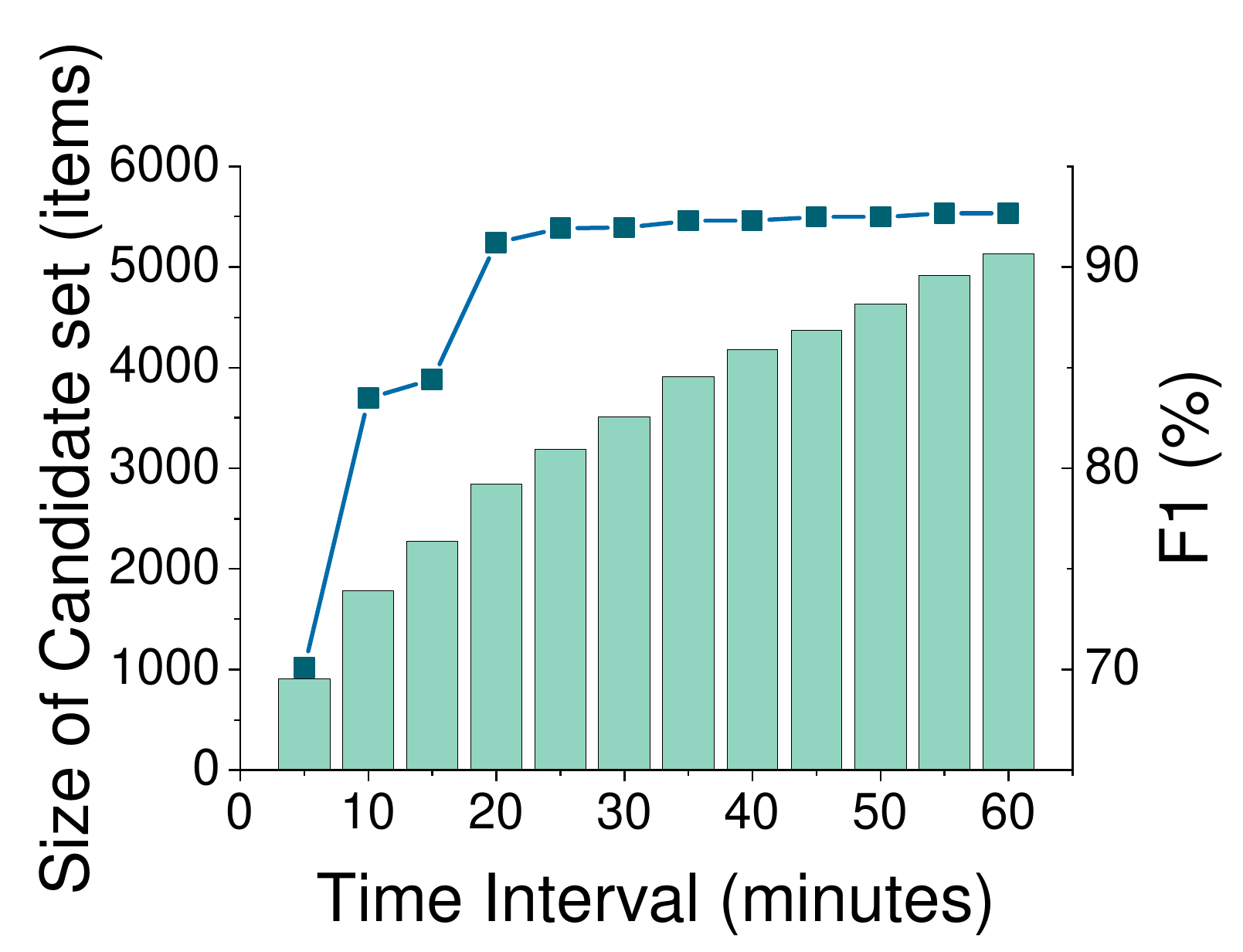}
	}
    \caption{The effect of time interval $\Delta$ on F1 and size of candidate set (Ethereum $\Rightarrow$ Polygon). Inflection points of $\Delta$ in Celer cBridge, Poly Network, and Multichain are 40, 140, and 55, respectively.}
    \label{fig:time_interval}
\end{figure}

\noindent\textbf{Time Interval Setting:} As previously mentioned, in the \textbf{M2}, we need to determine an optimal time interval for constraining the block range to be crawled. Typically, this time interval is set to 30 minutes based on the business rules of cross-chain bridges. However, to determine the optimal time interval constraint in real-world scenarios, we crawl candidate target transaction sets under various time interval settings. We record the corresponding set sizes and traceability effects. For instance, Fig.~\ref{fig:time_interval} illustrates the variation of candidate set sizes and traceability effects with time intervals for Celer cBridge, Multichain, and Poly Network. As the time interval increases, both the candidate set size and traceability effect exhibit an upward trend. The growth rate of candidate set size and traceability effect is more pronounced when the time interval is small, but subsequently levels off. When further increasing the time interval no longer yields significant improvements in the traceability effect, we can reasonably set the time interval by identifying the inflection point of the growth curve, thereby striking a balance between optimized traceability performance and computational efficiency. Note that we adopt the value of the inflection point as the optimal time interval $\Delta$, which will be used as the parameter setting for subsequent experiments.

\noindent\textbf{Evaluation Setting:} Based on the optimal time interval setting, we conducted 120 rounds of model training. The evaluation metrics employed include precision rate, recall rate, accuracy rate, and F1 score. Notably, we do not adopt the match rate (MR), zero hit rate (ZHR), and wrong hit rate (WHR) metrics mentioned in \contrastiveModel~\cite{lin2024connector}, as our correlation method is grounded in sorting rather than rule-based filtering, thereby eliminating the issues of zero hits and multiple hits.  

Table~\ref{tab:tracing_results} presents the results of our cross-chain transaction tracing experiments. Our proposed \myModel~demonstrates superior comprehensive capabilities in cross-chain transaction tracing tasks compared to rule-based methods like \contrastiveModel. Three key advancements are highlighted: 1) Complete task coverage: \myModel~successfully supports forward tracing (94.92\% F1), backward tracing (89.58\% F1), and bidirectional tracing (91.75\% F1), whereas \contrastiveModel~ - relying on predefined rule libraries—only achieves unidirectional tracing capability (TABLE~\ref{tab:tracing_results}). This addresses a critical limitation in existing solutions for multi-directional traceability. 
2) Automated feature interpretation: While CONNECTOR achieves 95.92\% accuracy in forward tracing through manual rule engineering, its performance hinges on pre-established cross-chain sample rules. In contrast, \myModel~achieves end-to-end relational reasoning through deep semantic modeling and can automatically extract the associated features. This is particularly advantageous for emerging cross-chain protocols (e.g., privacy-enhanced bridges), where labeled data for rule development is often scarce—a common challenge in anti-money laundering (AML) compliance scenarios targeting novel illicit activities. 
3) Architectural flexibility: When utilizing CodeBERT as the pre-training model, \myModel~achieves 91.75\% accuracy in bidirectional tracing. Although this slightly lags behind CONNECTOR’s unidirectional accuracy (95.92\%), its multi-task optimization framework breaks the conventional single-task paradigm. 


Furthermore, we conduct an in-depth analysis of the explicit traceability cues associated with each cross-chain bridge, revealing that the descriptions of chain entities in the majority of cross-chain event logs typically contain keywords such as \texttt{chain} and \texttt{id}. More specifically, entities on the source chain are predominantly denoted by terms with analogous meanings, including \texttt{from}, \texttt{source}, and \texttt{sender}, whereas entities on the target chain are primarily identified by terms such as \texttt{to}, \texttt{destination}, and \texttt{receipt}. The high-frequency occurrence of these keywords across different cross-chain bridge logs provides a robust foundation for the extraction of explicit traceability cues.

\begin{table*}[tbph]
\setlength{\abovecaptionskip}{0.cm}
  \centering
  \caption{Evaluating the effectiveness of \myModel~and existing methods in cross-chain transaction tracing. ``-" indicates that no experimental results can be obtained for this task by this method. Bold indicates the best result under this task, and italic indicates the second-best result.}
  \scalebox{0.9}{
    \begin{tabular}{m{3.6cm}|cccc|cccc|cccc}

    \toprule
    \multirow{2}[2]{*}{\scalebox{1.0}{\textbf{Model}}} & 
    \multicolumn{4}{c|}{\scalebox{1.0}{\textbf{Forward}}} &
    \multicolumn{4}{c|}{\scalebox{1.0}{\textbf{Backword}}} &
    \multicolumn{4}{c}{\scalebox{1.0}{\textbf{Bidirectional}}}
    \\
      & \scalebox{1.0}{\textbf{Pre}} & \scalebox{1.0}{\textbf{Rec}} & \scalebox{1.0}{\textbf{Acc}} & \scalebox{1.0}{\textbf{F1}} 
      & \scalebox{1.0}{\textbf{Pre}} & \scalebox{1.0}{\textbf{Rec}} & \scalebox{1.0}{\textbf{Acc}} & \scalebox{1.0}{\textbf{F1}}
      & \scalebox{1.0}{\textbf{Pre}} & \scalebox{1.0}{\textbf{Rec}} & \scalebox{1.0}{\textbf{Acc}} & \scalebox{1.0}{\textbf{F1}}
    \\
    \midrule
    \midrule
    \scalebox{1.0}{\contrastiveModel~} 
          & \scalebox{1.0}{\textbf{95.92\%}} & \scalebox{1.0}{\textbf{100\%}} & \scalebox{1.0}{\textbf{95.92\%}} & \scalebox{1.0}{\textbf{97.92\%}} 
          & \scalebox{1.0}{-} & \scalebox{1.0}{-} & \scalebox{1.0}{-} & \scalebox{1.0}{-}
          & \scalebox{1.0}{-} & \scalebox{1.0}{-} & \scalebox{1.0}{-} & \scalebox{1.0}{-}\\
    \midrule
    \scalebox{1.0}{\myModel~(Codebert)} 
          & \scalebox{1.0}{\textit{95.46\%}} & \scalebox{1.0}{\textit{94.64\%}} & \scalebox{1.0}{\textit{94.64\%}} & \scalebox{1.0}{\textit{94.92\%}} 
          & \scalebox{1.0}{\textbf{90.57\%}} & \scalebox{1.0}{\textbf{89.13\%}} & \scalebox{1.0}{\textbf{89.13\%}} & \scalebox{1.0}{\textbf{89.58\%}}
          & \scalebox{1.0}{\textbf{92.31\%}} & \scalebox{1.0}{\textbf{91.51\%}} & \scalebox{1.0}{\textbf{91.51\%}} & \scalebox{1.0}{\textbf{91.75\%}}\\
    \scalebox{1.0}{\myModel~(Graphcodebert)} 
          & \scalebox{1.0}{95.37\%} & \scalebox{1.0}{94.57\%} & \scalebox{1.0}{94.57\%} & \scalebox{1.0}{94.84\%} 
          & \scalebox{1.0}{90.22\%} & \scalebox{1.0}{88.55\%} & \scalebox{1.0}{88.55\%} & \scalebox{1.0}{89.04\%}
          & \scalebox{1.0}{91.10\%} & \scalebox{1.0}{89.77\%} & \scalebox{1.0}{89.77\%} & \scalebox{1.0}{90.06\%}\\
    \scalebox{1.0}{\myModel~(Unixcode)} 
          & \scalebox{1.0}{94.93\%} & \scalebox{1.0}{93.99\%} & \scalebox{1.0}{93.99\%} & \scalebox{1.0}{94.30\%} 
          & \scalebox{1.0}{\textit{90.36\%}} & \scalebox{1.0}{\textit{88.71\%}} & \scalebox{1.0}{\textit{88.71\%}} & \scalebox{1.0}{\textit{89.20\%}}
          & \scalebox{1.0}{91.40\%} & \scalebox{1.0}{90.47\%} & \scalebox{1.0}{90.47\%} & \scalebox{1.0}{90.70\%}\\
    \scalebox{1.0}{\myModel~(CodeT5)} 
          & \scalebox{1.0}{95.03\%} & \scalebox{1.0}{94.17\%} & \scalebox{1.0}{94.17\%} & \scalebox{1.0}{94.45\%} 
          & \scalebox{1.0}{87.41\%} & \scalebox{1.0}{83.59\%} & \scalebox{1.0}{83.59\%} & \scalebox{1.0}{84.39\%}
          & \scalebox{1.0}{\textit{92.06\%}} & \scalebox{1.0}{\textit{91.32\%}} & \scalebox{1.0}{\textit{91.32\%}} & \scalebox{1.0}{\textit{91.53\%}}\\
    \bottomrule    
    \end{tabular}%
  }
  \label{tab:tracing_results}%
\end{table*}%

\begin{figure}[h]
\setlength{\abovecaptionskip}{0.2cm}
    \centering
    \includegraphics[width=0.8\linewidth]{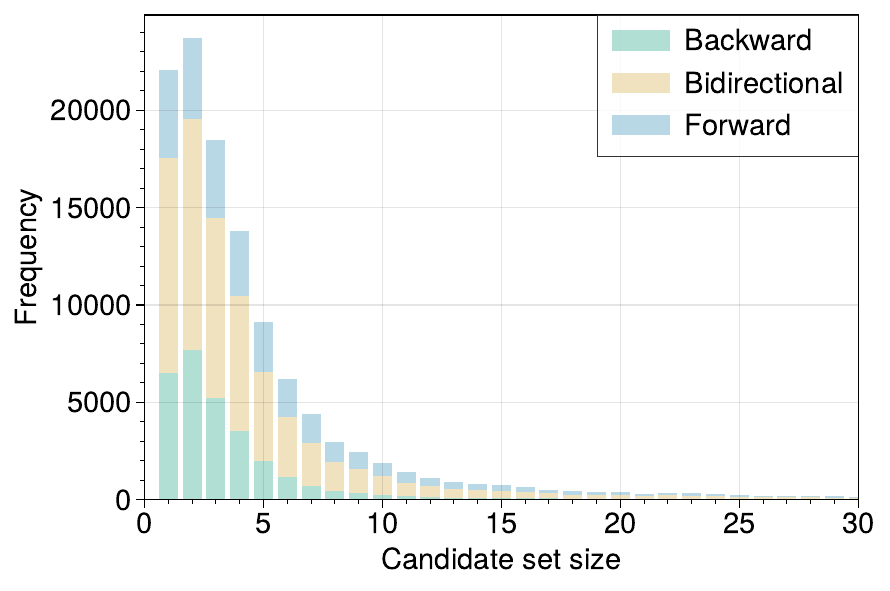}
    \caption{Distribution of the size of candidate transaction sets. The frequency of candidate transaction sets decreases with increasing transaction set size for all three tracing. Most query transactions correspond to candidate transaction sets with sizes less than 10.}
    \label{fig:frequency}
\end{figure}

Additionally, we examine the contribution of \textbf{M2} to the cross-chain transaction tracing process, focusing on the distribution of the number of candidate transaction sets generated following \textbf{M2} recall. As illustrated in Fig.~\ref{fig:frequency}, it is evident that the frequency of candidate transaction sets decreases as their size increases, with the majority of queried transactions corresponding to candidate sets of fewer than 10 transactions. Notably, \textbf{M2} significantly diminishes the search space of candidate transactions that require processing in the tracing process, thereby enhancing retrieval efficiency, compared to directly crawling block data adjacent to the transaction time on the target chain.

\subsubsection{Capability of Automatic Learning and Generalization}
To further validate the automated learning capabilities of \myModel, we select CodeBert~\cite{feng2020codebert}, which excels in transaction tracing, as the pre-trained model. Since \contrastiveModel~\cite{lin2024connector} only supports forward tracing, to ensure the fairness of the experiment, we will focus solely on evaluating performance of \myModel~in forward tracing. Additionally, because \contrastiveModel~\cite{lin2024connector} relies on predefined rules and cannot perform association operations when dealing with transactions involving unknown cross-chain bridges, we will assess \myModel~independently.

\noindent\textbf{Evaluation Setting:} In the experimental design, we establish three different training set sizes: {25\%}, {50\%}, and {75\%}. Specifically, we divide the real-labeled dataset of cross-chain transactions from 12 cross-chain bridges. For the {25\%} group, we use the cross-chain transactions from Polybridge, Stargate, Debridge, and Router protocol as the training set. At the same time, the remaining bridges are split equally into the validation and test sets (the same approach is used for other groups). In the {50\%} group, we select cross-chain transactions from Multichain, Transit swap, Stargate, Synapse Protocol, Connext Bridge, Symbiosis, Debridge, and Router Protocol for the training set. Finally, in the {75\%} group, we include cross-chain transactions from Multichain, Celer cbridge, Transit Swap, stargate, Synapse Protocol, Symbiosis, Debridge, and Router Protocol. We conduct tests on cross-chain bridges not included in the training process to assess the ability of \myModel~to learn automatically in unseen cross-chain environments.

\begin{figure}[h]
\setlength{\abovecaptionskip}{0.2cm}
    \centering
    \includegraphics[width=0.8\linewidth]{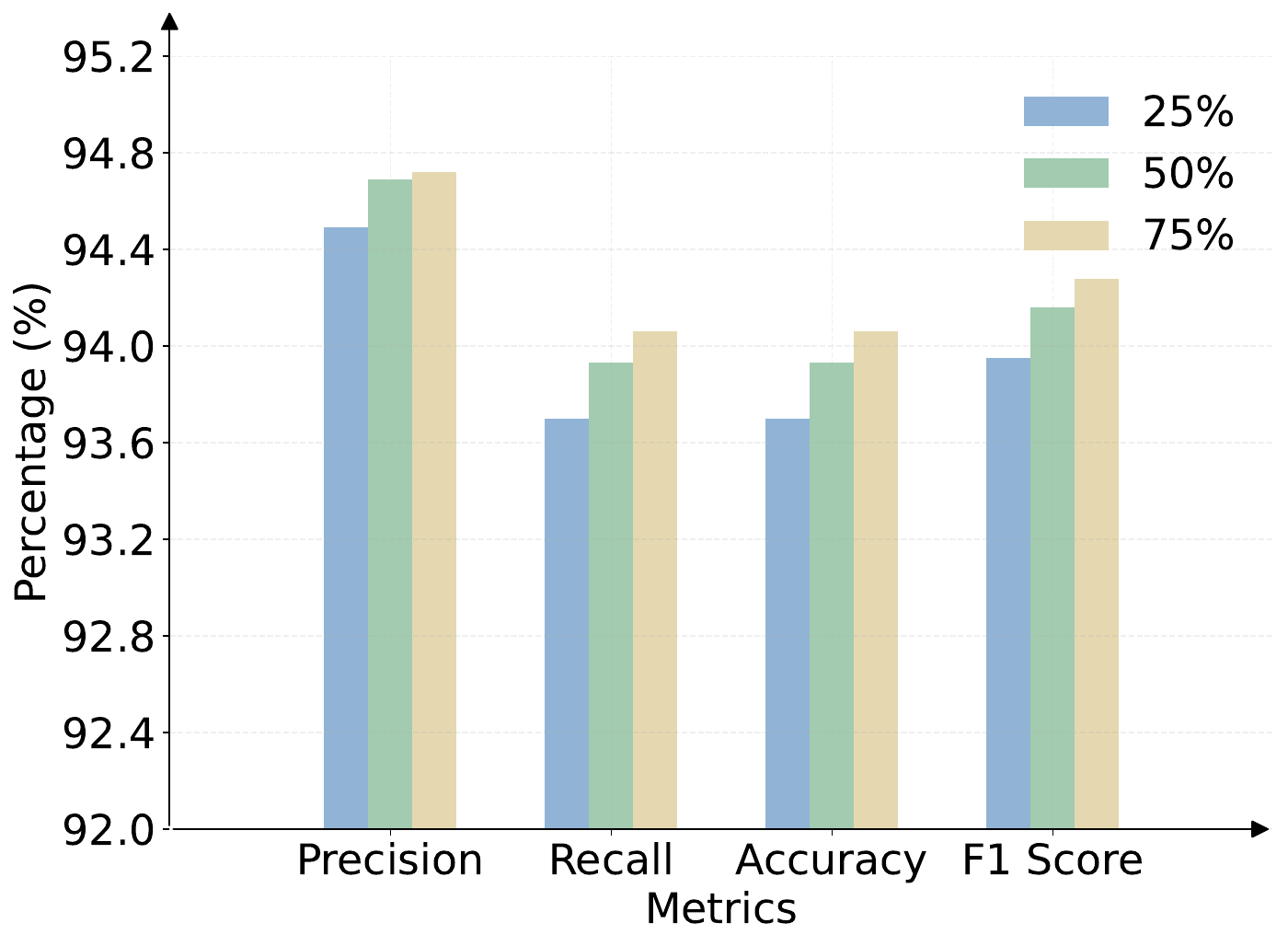}
    \caption{The performance indicator values are above {93\%} for all three groups. As more cross-chain bridges appear in the training set, the indicator values are higher, and the results are less affected by the unseen cross-chain bridges.}
    \label{fig:auto}
\end{figure}

As depicted in Fig.~\ref{fig:auto}, the experimental results illustrate the performance of \myModel~across three different configurations. By examining the trends in four key metrics—Precision, Recall , Accuracy , and F1 Score. It is evident that the tracing performance of \myModel~enhances as the number of learned cross-chain bridges increases. Notably, even in the experimental group with only {25\%} pre-trained cross-chain bridge objects, \myModel~achieves an F1 score of {93.95\%}, comparable to the other two groups. This demonstrates the strong automatic learning capabilities and robustness of \myModel, indicating that its performance is less dependent on the dataset size, making it highly adaptable.

\subsection{RQ2: Application to attack transactions}
To address RQ2, we employ \myModel~to trace and analyze real-world cross-chain attack transactions. Cross-chain attacks can be broadly classified into three categories: source chain attacks, target chain attacks, and cross-chain process attack~\cite{wu2024safeguarding}. Notably, cross-chain process attack are primarily facilitated through cross-chain transactions, it relies on the collaboration of transactions on both chains to complete the attack, and is a high-level attack mode unique to cross-chain scenarios. This attack is unique to cross-chain scenarios. In contrast, the other two types of attacks do not directly involve cross-chain transactions. Consequently, we trace the cross-chain process attack transactions provided by \cite{wu2024safeguarding} and successfully identify 20 cross-chain transaction pairs. Among these, 18 transactions from Ethereum (ETH) to Binance Smart Chain (BSC) originated from the Qubit cross-chain bridge, 1 cross-chain transaction from BSC to ETH was attributed to the meterio cross-chain bridge, and 1 cross-chain transaction from ETH to BSC is linked to the Poly Network cross-chain bridge. However, since the Qubit cross-chain bridge and the meterio cross-chain bridge do not meet our criteria for selecting experimental cross-chain bridges, such as the absence of a query function for cross-chain transaction pairs, they are not included in the experimental scope of RQ1. Instead, they were utilized for exploratory purposes in additional cross-chain attack traceability experiments.

Through a comprehensive analysis of these 20 cross-chain attack transactions, we draw the following conclusions: 1) In the attacked Qubit cross-chain bridge transactions, the actual deposit amounts are zero, whereas the actual withdrawal amounts exist. This anomaly is attributed to a vulnerability in the Qubit bridge contract, which the attacker exploited to fabricate fictitious deposits; 2) In the attacked Meterio cross-chain bridge transaction, the actual deposit amount is zero. This is because the cross-chain tokens were utilized as encapsulated native tokens that were not burned or locked. However, the native token had been unencapsulated during the cross-chain operation, and the funds had been transferred to the handler contract; 3) In the attacked Poly Network cross-chain bridge transaction, the withdrawal amount is substantially higher than the deposit amount. This discrepancy is due to the off-chain repeater being replaced by a validator by the attacker, resulting in the tampering of the withdrawal transaction amount. In summary, when the deposit amount of a cross-chain transaction exceeds the withdrawal amount, or the cross-chain fee percentage surpasses {3\%}, the likelihood of anomalies in cross-chain transactions increases. This discovery provides a robust foundation for cross-chain transaction anomaly detection.

\subsection{RQ3: Application to money laundering transactions}
\begin{figure}[h]
\setlength{\abovecaptionskip}{0.2cm}
    \centering
    \includegraphics[width=1\linewidth]{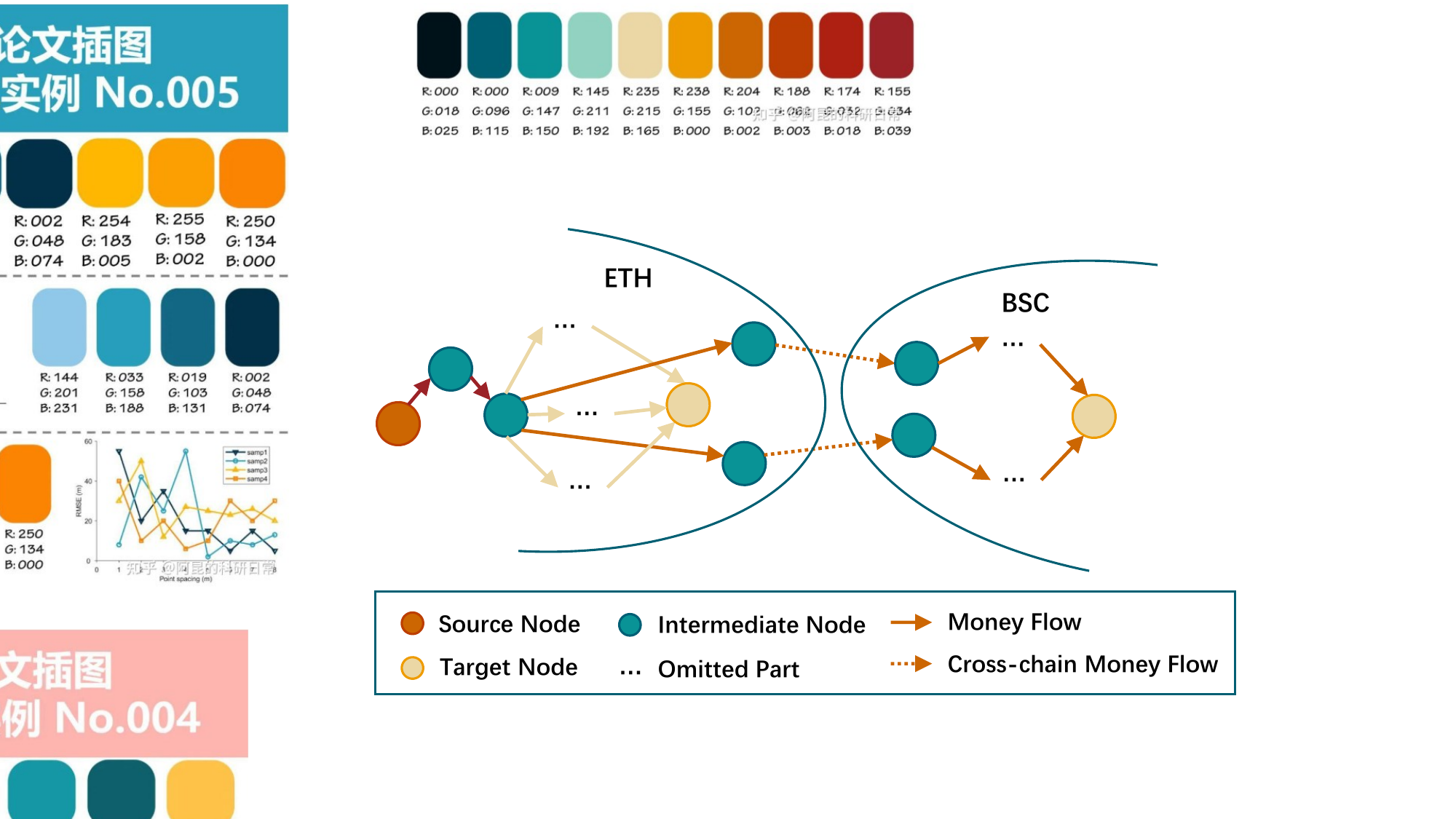}
    \caption{Visual representation of cross-chain money laundering flows derived from the Arthur0xWalletHacker dataset. The color of the money flow represents the proportion of the laundering amounts, where the money flow on ETH is lighter and the money flow to the BSC chain is darker. This indicates that in this example, the proportion of the laundering amounts across chains is high.}
    \label{fig:rq3}
\end{figure}

To address RQ3, we employ \myModel~to trace and analyze real-world cases of cross-chain money laundering transactions. Utilizing the Ethereum money laundering case data provided by \cite{wu2023towards}, we successfully identified and traced 10 cross-chain transactions associated with money laundering activities. The distribution of these transactions is as follows: three from the Coinrail case, one from the KucoinHacker case, five from the Arthur0xWalletHacker case, and one from the UpbitHack case. To illustrate the operational flow of cross-chain money laundering, we conducted a visual analysis of the laundering pathways involved. For instance, the specific route of cross-chain money laundering derived from the Arthur0xWalletHacker dataset is depicted in Fig.~\ref{fig:rq3}.

The source node represents the original hacker account involved in this incident, while the intermediary nodes denote the layered accounts through which the money laundering occurs. The target node corresponds to the hacker account where the laundered funds ultimately converge. The color intensity of the solid line arrows indicates the proportion of the laundering amounts throughout the entire process; darker colors signify a higher proportion, while lighter colors indicate a lower proportion. The dashed arrows represent the proportions of funds involved in cross-chain money laundering.

We observe that the cross-chain money laundering process exhibits certain similarities to the single-chain money laundering process, as both undergo three stages: placement, layering, and aggregation~\cite{wu2023towards}. However, unlike single-chain money laundering, the cross-chain variant introduces cross-chain operations during the layering stage. This approach disperses the flow of funds across two blockchains, significantly increasing the complexity of the laundering pathways and presenting greater challenges for asset tracking.

\section{Related Work}
\label{subsec:related_work}

In this section, we briefly review existing research work on transaction tracing in single-chain scenarios and cross-chain scenarios.

\subsection{Single-chain Transaction Tracing}
Existing methods for tracing single-chain transactions are primarily designed to address specific security issues. These approaches rely on predefined heuristic rules and account classifications to track the flow of funds, ultimately aiming for anomaly detection or anti-money laundering (AML) objectives. \cite{chen2019market,gao2020tracking,wu2021towards,wu2023know,xia2021trade} on blockchain anomaly analysis and detection often emphasizes account classification, employing specific patterns and rules to trace the origins of illicit transactions. Examples include phishing account detection~\cite{li2022ttagn,wu2020phishers}, Ponzi scheme identification~\cite{chen2018detecting}, and fraudulent account classification~\cite{hu2023bert4eth,wang2022demystifying}. In the Bitcoin ecosystem, AML research typically utilizes Elliptic datasets~\cite{weber2019anti} to classify accounts, distinguishing between illegal and legitimate entities. Conversely, in the Ethereum space, AML initiatives use heuristic rules to identify potential money laundering networks~\cite{wu2023towards}. However, all these methods are tailored to a single ledger and are not readily applicable to cross-chain scenarios.

\subsection{Cross-chain Transaction Tracing}
Existing methods for cross-chain transaction tracing primarily focus on the study of CeFi bridges. Yousaf et al.~\cite{Yousaf2019Tracing} is the first to explore the correlation of cross-chain transactions in externally owned account (EOA)-based CeFi bridges. This work proposes a method for associating cross-chain transactions based on a heuristic rule-matching algorithm and analyzes the patterns of these transactions. Concurrently, \cite{Zhang2022CLTracer} developes a cross-chain transaction clustering framework known as CLTracer, which leverages address relationships. This framework successfully detects cross-chain transaction behaviors by integrating cross-chain clustering with heuristic algorithms. However, these methods have notable research limitations: 1) Low accuracy: The original tracing algorithm within~\cite{Yousaf2019Tracing} relies on a centralized bridge API to identify deposit transactions on the blockchain, achieving only {80\%} accuracy, indicating significant room for improvement; 2) Non-independence: Existing methods depend on the internal APIs of CeFi cross-chain bridges to match withdrawal transactions, making them unsuitable for direct application in decentralized finance (DeFi) cross-chain scenarios.

Additionally, our team has introduced a novel cross-chain transaction tracing tool called \contrastiveModel~\cite{lin2024connector}, specifically designed for DeFi bridges. As previously mentioned, \contrastiveModel~currently supports only one-way tracing from the source chain to the destination chain and is constrained by predefined rules, resulting in limited automation in the transaction correlation process. These factors hinder \contrastiveModel~'s effectiveness in tracing complex cross-chain transaction attacks and money laundering scenarios.

\section{Conclusion}
\label{subsec:conclusion}
In this paper, we propose the first automated, bi-directional transaction traceability tool oriented towards a contract-based DeFi cross-chain bridge design, referred to as \myModel. This tool enhances the identification and association of cross-chain transactions through contract log mining. It begins with semantic extraction to identify cross-chain transactions, followed by using named entity recognition to automatically extract explicit cross-chain cues for candidate crawling. Subsequently, information retrieval techniques are employed to encode implicit cross-chain clues, facilitating precise association and effective tracing of cross-chain transactions. Additionally, we conducted extensive experiments on a real-world cross-chain transaction dataset sourced from various cross-chain bridge applications. The results demonstrate that \myModel~achieves an F1 score of {94.92\%} in forward tracing, {89.58\%} in backward tracing, and {91.75\%} in bi-directional tracing. In terms of practical application, we leverage \myModel~to trace and analyze real-world cross-chain attack transactions and money laundering cases, successfully identifying 20 cross-chain attack transactions and 10 flows associated with cross-chain money laundering. In summary, \myModel~significantly enhances the traceability and security of bridges within the multi-chain DeFi ecosystem, thereby establishing a solid foundation for tasks such as cross-chain attack transaction detection and cross-chain anti-money laundering.

\bibliographystyle{IEEEtran}
\bibliography{ref}


\begin{IEEEbiography}
    [{\includegraphics[width=1in,height=1.25in,clip,keepaspectratio]{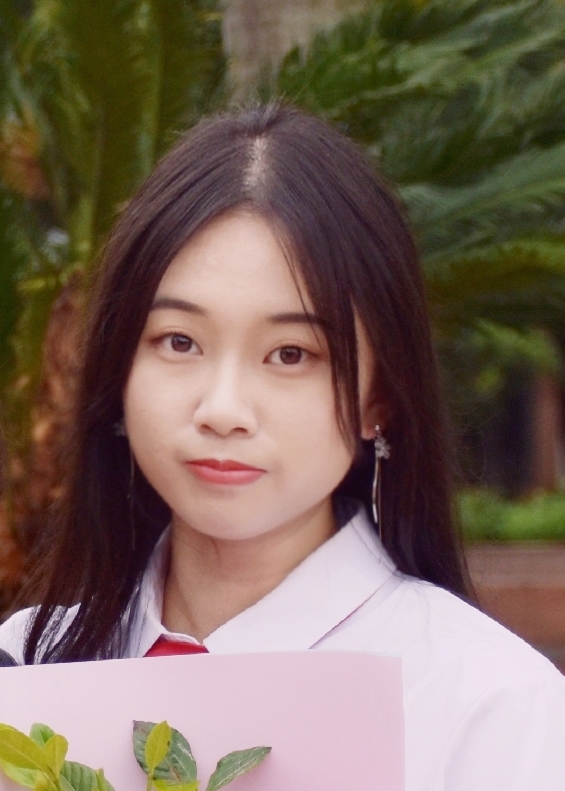}}]{Dan Lin} 
     (Member, IEEE) received her Ph.D. degree in software engineering with Sun Yat-sen University, China, in 2024. She is currently a postdoctoral researcher with the School of Software Engineering, Sun Yat-sen University, Zhuhai, China. She is also an Executive Committee Member of the Special Committee on Service Computing of the China Computer Federation (CCF TCSC). Her current research interests include blockchain, cross-chain technology, anti-money laundering, applications of network science.
\end{IEEEbiography}

\begin{IEEEbiography}
	[{\includegraphics[width=1in,height=1.25in,clip,keepaspectratio]{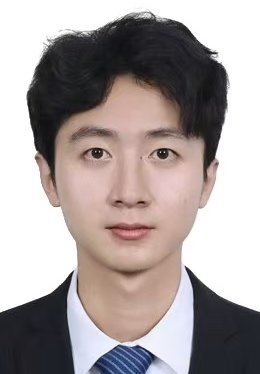}}]{Ziye Zheng} received his B.Eng. degree in software engineering from South China Normal University, Foshan, China, in 2023. He is currently studying toward the M.Sc. degree with the School of Computer Science and Engineering, Sun Yat-sen University, Guangzhou. His current research interests include blockchain transaction analysis, cross-chain transaction tracing.
\end{IEEEbiography}

\begin{IEEEbiography} 
	[{\includegraphics[width=1in,height=1.25in,clip,keepaspectratio]{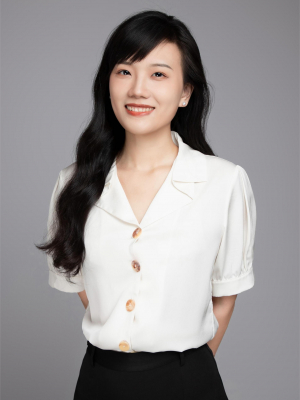}}]{Jiajing Wu} (Senior Member, IEEE) received the B.Eng. degree in communication engineering from Beijing Jiaotong University, Beijing, China, in 2010, and the Ph.D. degree from Hong Kong Polytechnic University, Hong Kong, in 2014. She was awarded the Hong Kong Ph.D. Fellowship Scheme during her Ph.D. study in Hong Kong (2010--2014). 
 
    She is currently an Associate Professor with the School of Software Engineering, Sun Yat-sen University, Zhuhai, China. Her research focus includes blockchain, graph mining, network science. She serves as an Associate Editor for {\sc IEEE Transactions on Circuits and Systems II: Express Briefs.}
\end{IEEEbiography}

\begin{IEEEbiography}
    [{\includegraphics[width=1in,height=1.25in,clip,keepaspectratio]{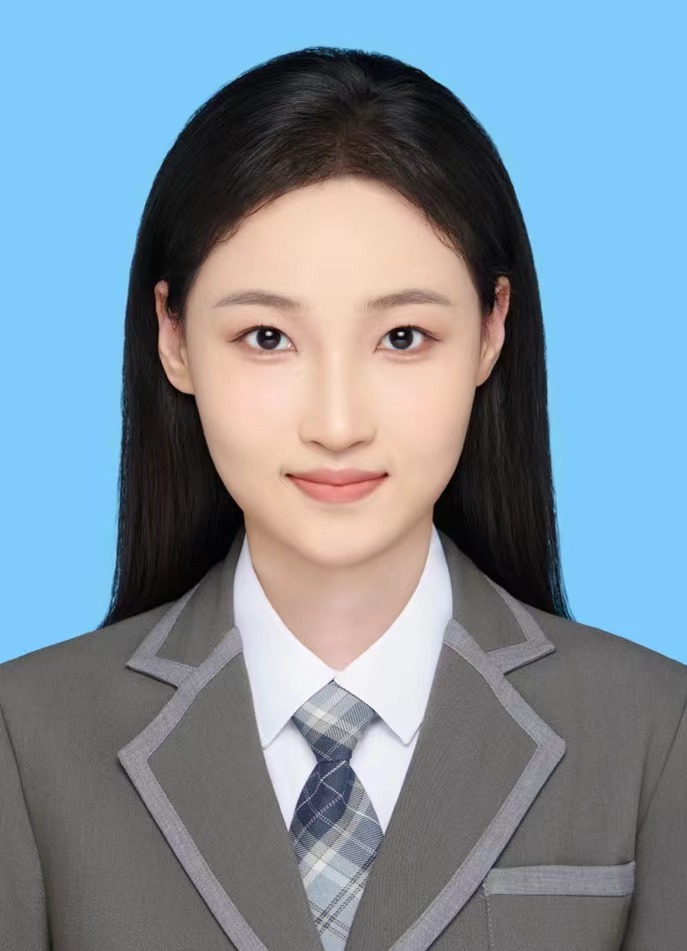}}]{Jingjing Yang} received the B.Eng. degree in computer science and technology from the Ocean University of China, Qingdao, China, in 2023. She is currently studying toward the M.Sc. degree with the School of Computer Science and Engineering, Sun Yat-sen University. Her current research interests include NFT, Web3, risk management, and phishing scam detection.
\end{IEEEbiography}

\begin{IEEEbiography}
    [{\includegraphics[width=1in,height=1.25in,clip,keepaspectratio]{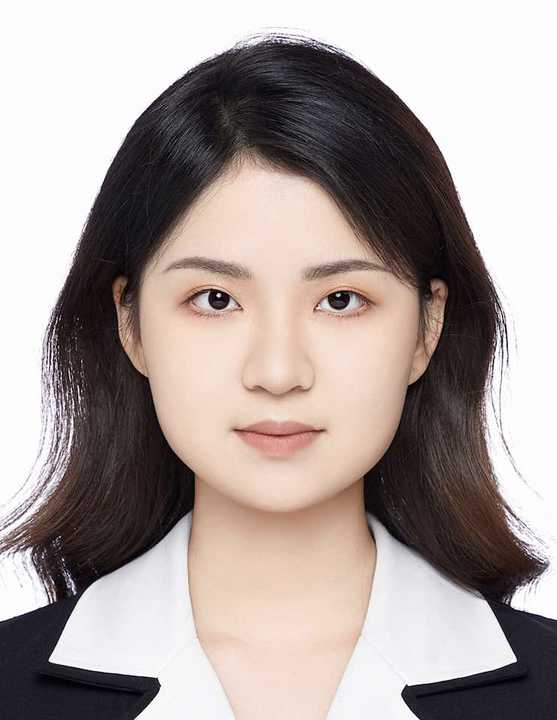}}]{Kaixin Lin} received the B.Eng. degree from the School of Computer Science, South China Normal University, Guangzhou, China, in 2022. She is working toward the M.Sc. degree with the School of Computer Science and Engineering, Sun Yat-sen University, Guangzhou, China. Her research interests include blockchain, cross-chain bridge, and graph mining.
\end{IEEEbiography}

\begin{IEEEbiography}
    [{\includegraphics[width=1in,height=1.25in,clip,keepaspectratio]{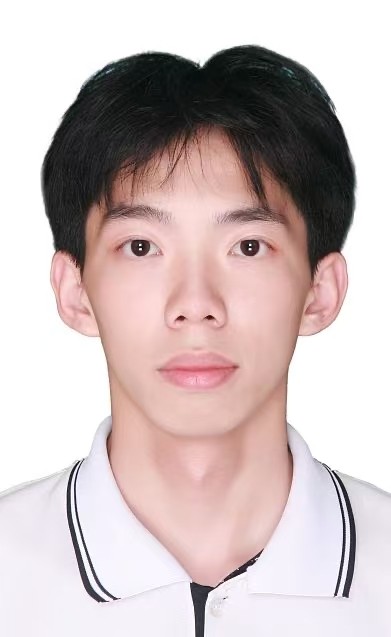}}]{Huan Xiao} is currently pursuing his Bachelor of Engineering degree in the School of Artificial Intelligence, Guizhou University, Guiyang, China, with an expected graduation in 2025. He has been admitted to the School of Software Engineering, Sun Yat-sen University, Zhuhai, China. His research interests include blockchain transaction analysis and cross-chain bridge analysis.
\end{IEEEbiography}

\begin{IEEEbiography}
	[{\includegraphics[width=1in,height=1.25in,clip,keepaspectratio]{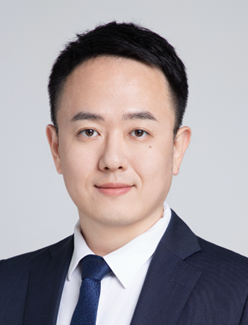}}]{Bowen Song} received his Ph.D. degree in Applied Math and Statistics from Stony Brook University in 2015.  He joined Ant Group in 2017 and now serves as a senior staff algorithm engineer in the anti-money-laundering algorithm team.  His research interests mainly focus on the area of behavior sequential learning, deep graph learning, and their applications in financial risk management and web3.
\end{IEEEbiography}

\begin{IEEEbiography}
	[{\includegraphics[width=1in,height=1.25in,clip,keepaspectratio]{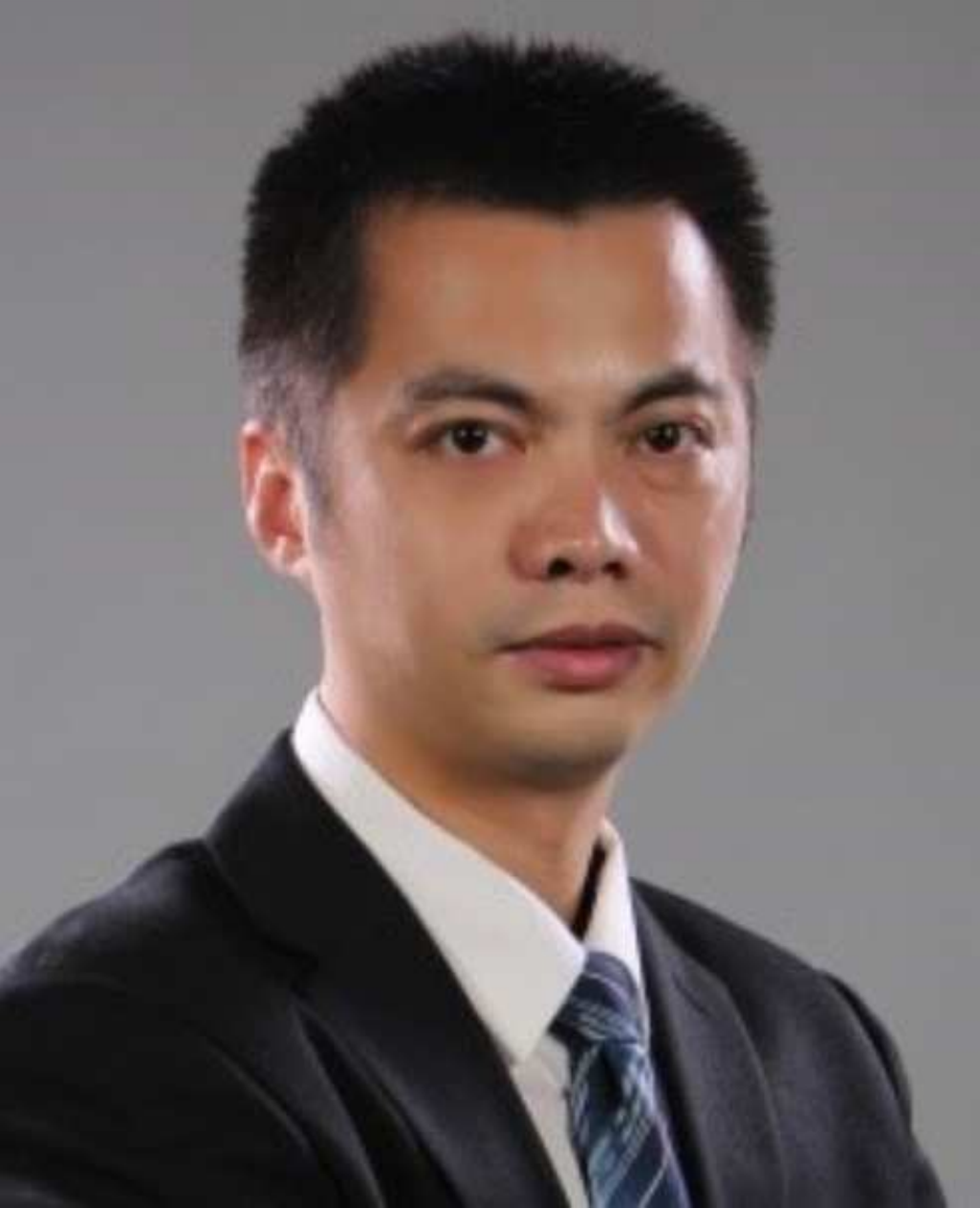}}]{Zibin Zheng}(Fellow, IEEE) is currently a Professor and the Deputy Dean with the School of Software Engineering, Sun Yat-sen University, Guangzhou, China. He authored or coauthored more than 200 international journal and conference papers, including one ESI hot paper and ten ESI highly cited papers. According to Google Scholar, his papers have more than 28,000 citations. His research interests include blockchain, software engineering, and services computing. He was the BlockSys’19 and CollaborateCom16 General Co-Chair, SC2’19, ICIOT18 and IoV14 PC Co-Chair. He is a Fellow of the IET. He received several awards, including the Top 50 Influential Papers in Blockchain of 2018, the ACM SIGSOFT Distinguished Paper Award at ICSE2010, the Best Student Paper Award at ICWS2010.
\end{IEEEbiography}

\vfill

\end{document}